\documentclass[final,3p,times,twocolumn,authoryear]{elsarticle}

\usepackage[utf8]{inputenc}
\usepackage{enumitem}
\usepackage{amssymb}
\usepackage{amsmath,bm}
\usepackage{graphicx}
\usepackage{subfig}
\usepackage{multirow}
\usepackage[normalem]{ulem}
\usepackage{appendix}
\usepackage{changepage,geometry}  
\usepackage{multicol}
\usepackage{natbib} 
\usepackage[dvipsnames]{xcolor}
\usepackage{hyperref} 
\hypersetup{hidelinks} 

\journal{Icarus}

\begin{document}

\begin{frontmatter}



\title{Ice Deposition Fronts In Porous Bodies From Transient Heating Events In a Protoplanetary Disk} 


\author[a1]{Stephen Li\corref{cor1}}
\ead{sluniews@ur.rochester.edu}
\author[a1]{Alice C. Quillen}
\author[a1]{Adam E. Rubinstein}
\author[a1]{Dominique Segura-Cox}
\author[a2]{Kevin Righter}

\address[a1]{Department of Physics and Astronomy, University of Rochester, Rochester, NY 14627, USA}
\address[a2]{Department of Earth and Environmental Science, University of Rochester, Rochester, NY 14627, USA}
 \cortext[cor1]{Corresponding author}
\begin{abstract}

Using a 1D mass and heat transport model, we numerically integrate heat flow and gas transport in a porous body exposed to a transient heating event while embedded in a protoplanetary disk. When small icy grains are heated, volatiles sublimate, enriching the disk with volatile gases. When a porous body enters this heated, volatile-rich environment, volatile gases diffuse throughout the cool, porous body and deposit ice where the partial pressure of a volatile exceeds its vapor pressure. 
We simulate sublimation and deposition fronts of water, carbon dioxide, and carbon monoxide. Our simulations show that an ice deposition front forms and moves deeper into the porous body as the body warms. The amount of nebular gas deposited in an initially dry body is usually extremely low; however, in an initially icy body, an ice deposition front contains locally sublimated volatiles. In this case, the front can increase the ice volume fraction (by a factor of 2) in a thin layer below the surface. 
We find that the propagation speed, propagation strength, and final depth of an ice deposition front primarily depend on pore size. We propose that nebular heating events can alter the subsurface morphology and physical properties of porous icy objects embedded in a protoplanetary disk. 
\end{abstract}






\end{frontmatter}


\section{Introduction}

Chondrules are small (diameter 0.1 to 1 mm) igneous spheres that are commonly found within meteorites. The properties of chondrules imply that small rocky particles in the protoplanetary disk experienced transient heating events \citep{Brearley_1998}.
In this study, we explore how transient heating could have affected large porous bodies that could also have been present in the protoplanetary disk. 

Models for chondrule formation include nebular shocks \citep{Ciesla_2002, Desch_2002}, planetary bow shocks \citep{Tanaka_2013, Mann_2016}, impact jetting \citep{Sanders_2012, Cashion_2025} or vapor plumes caused by catastrophic impacts \citep{Stewart_2000}. 
Formation of chondrules is expected to occur in the disk midplane \citep{Morris_Boley_2018} and require cooling times ranging from tens of minutes to a few hours \citep{Fujii_1983, Stammler_2014}. 
The heat required for melting rocky material and forming chondrules needs to be quite high (temperature $T \gtrsim 1500$K; \citealt{Costa_2017}). However, lower-temperature heating events that are capable of sublimating volatile compounds, but not high enough to melt rocky material, could also occur in protoplanetary disks \citep{Podolak_2011} and may be more frequent or numerous than high temperature heating events \citep{Ormel_2008,Matsumoto_2023}. 

While chondrule formation likely took place within 2-3 AU of the Sun \citep{Iida_2001, Boss_2005}, shock heating caused by spiral density waves could have occurred throughout the protoplanetary disk. Models that are locally isothermal may poorly approximate the locations of ice lines \citep{Podolak_2011,Ziampras_2020}. 
Transient heating events can facilitate chemical reactions that would not have otherwise occurred; as inferred from fine grained phyllosilicate rims on chondrules \citep{Cyr_1998, Ciesla_2003}. 
Heating events in the disk could vary the thickness of the water ice layer on a small aggregate after cooling \citep{Sirono_2024}. 

 



Both asteroidal materials \citep{Britt_2002,Garvie_2024} and 
icy pebbles that are formed in a protoplanetary nebula are likely to be fluffy or porous 
\citep{Kataoka_2013,Lorek_2018,Zivithal_2025}. 
Comets too are porous \citep{Prialnik_1995,Thomas_2013, Fulle_2016,Kossacki_2018,Patzold_2019,ORourke_2020}. 
Sublimated comet material escapes into the interplanetary medium \citep{Whipple_1950, Gombosi_1986, Bouziani_2022}. However, some models predict that sublimated gas permeate the comet body and be deposited below the comet's surface where the temperature remains low \citep{Kossacki_2015,Spohn_2015,Gundlach_2018,Fellows_2020,Zivithal_2025}. A transient heating event that can melt a mm diameter grain may not last long enough to sublimate all the ice from the center of 10 cm diameter or larger porous body. 

\begin{figure}[!t]\centering
\includegraphics[width=3truein]{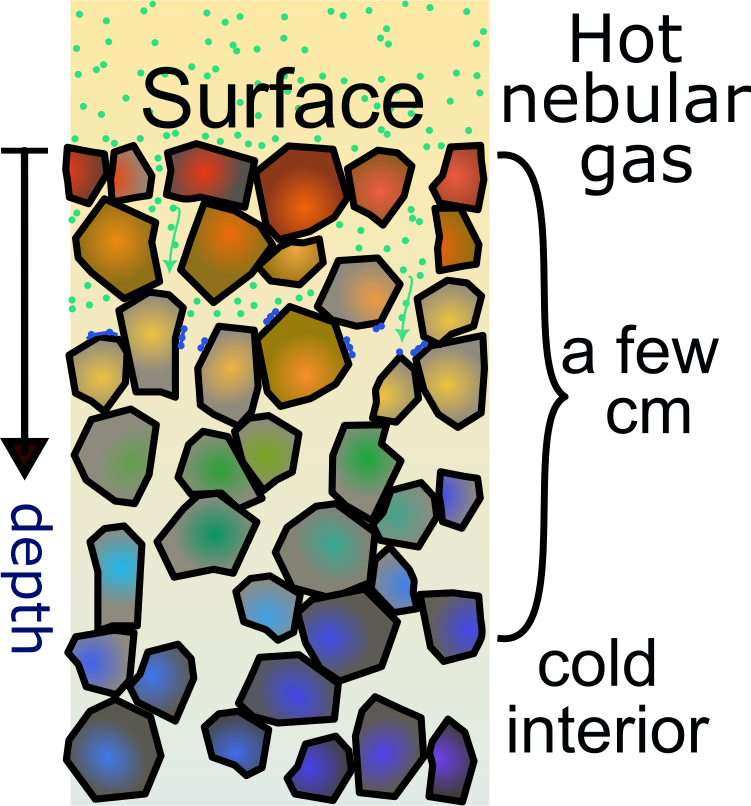}
\caption{A dry, porous and rocky body is heated at the surface by a hot nebula that contains volatile gases. Volatiles (represented as green dots) can penetrate the porous body and can be deposited as ices below the surface where it is cooler. The colors of the porous bodies illustrate a temperature gradient. The interior of the body acts like a cold trap. 
\label{fig:illust1}}
\end{figure}

\begin{figure}[!t]\centering
\includegraphics[width=3.2truein]{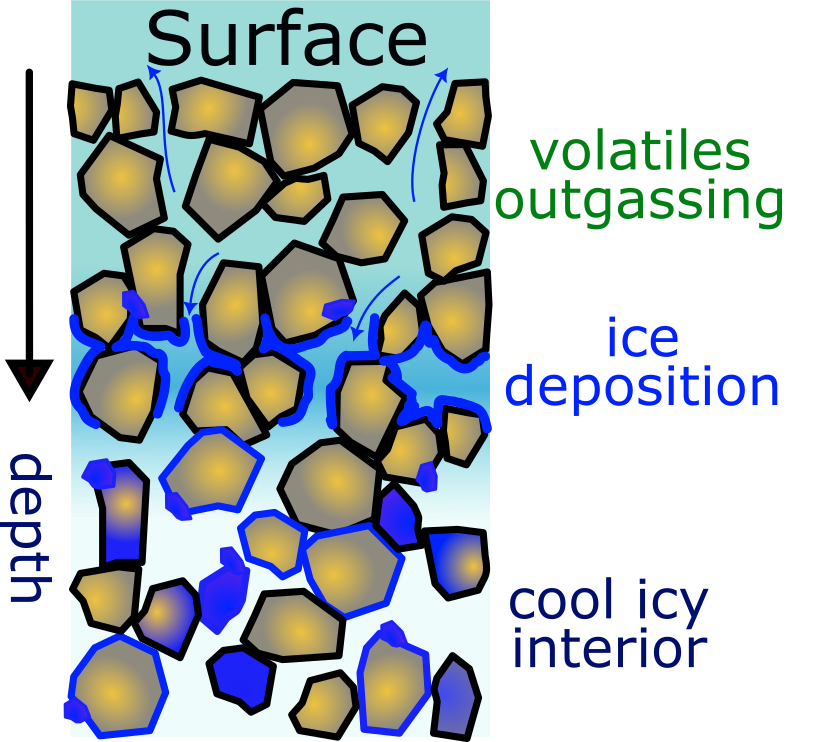}
\caption{An initially icy and porous body is heated at its surface due to a short duration heating event in the protoplanetary disk. Sublimating volatiles below the surface out-gas and also penetrate into the interior where they can be deposited as ices. Dark blue represents ice and aquamarine represents volatile rich gas. Both sublimating and ice deposition layers can advance into the body as the body is heated. After the nebula's temperature drops back to an equilibrium value, ice remains concentrated in a thin layer below the surface. 
\label{fig:illust2}}
\end{figure}

In this study, we investigate the effects of transient heating events on the internal ice fraction of porous bodies that could be present in a protoplanetary disk. Local heating in a disk can cause small (mm-sized) icy grains to rapidly sublimate, enriching the disk's volatile gas content. Consequently, we consider the possibility that the nebular gas is rich in warm volatiles while it is hot. Volatile gases present in the protoplanetary nebula could penetrate the porous body, and be deposited in the form of ices below the surface, as illustrated in Figure \ref{fig:illust1}. As an icy body's surface is heated, its sublimating volatiles would both out-gas into the disk environment and penetrate deeper within the porous body where they can be deposited as ice, as illustrated in Figure \ref{fig:illust2}. When heated, a porous medium can experience an advancing internal sublimation front where ice is sublimated (e.g. \citealt{Kulikovskii_2002,Scarpa_2002,Warning_2015,Bouziani_2022}).
If the interior of 
a porous body is cold, it could act like a cold trap causing an advancing internal condensation or deposition front (e.g., \citealt{Kulikovskii_2002,Scarpa_2002,Lei_2024}). In the context of comets, ice is deposited below the sublimation front in a layer called a `sinter layer' \citep{Kossacki_2015,Spohn_2015,Zivithal_2025}.

\section{Transport Equations for the Numerical Model}

The gas pressure in the midplane of a circumstellar disk is lower than that typical of a planetary atmosphere (see Figure 1 by \citealt{Quillen_2024}). Multiphase transport models that are developed for applications on Earth (e.g., \citealt{Warning_2015}) must be adapted to apply at the low pressures of a circumstellar disk. Inside a porous body that is embedded in the disk (and excluding extreme events such as a high-velocity impacts), the vapor pressure of most volatiles would be below the volatile's triple point (see \citealt{Fellows_2020}). For example, water's triple point is at a temperature of 273 K and 
at a pressure of 611 Pa, but the disk midplane pressure at 1 AU from the Sun is only a few Pa in a protosolar disk model similar to the minimum mass solar nebula \citep{Quillen_2024}. When the gas partial pressure of a molecule exceeds its vapor pressure, the gas phase is deposited on a surface as solid ice. We use the word `deposited' instead of `condensed' to make it clear that ice is deposited and that a liquid phase is unlikely to be present. 

The physical model we use for an icy porous body heated by hot ambient nebular gas is a modification of the icy conglomerate model for comets by \citet{Whipple_1950} (also see \citealt{Gombosi_1986,Bouziani_2022}). This model resembles the multiphase porous medium transport model by \cite{Warning_2015} for sublimation fronts that occur during vacuum freeze drying of beef. However, following \citet{Schweighart_2021} and as we describe below, because the protoplanetary disk gas has a low pressure, the mean free path is large compared to the pore size. This limit, known as the the Knudsen limit, lets us simplify the transport model. The surface boundary condition for the icy porous body differs from that of a comet as heating of the body is primarily due to a temperature increase in the gas phase of the circumstellar disk, rather than solar illumination. 

Within an icy conglomerate, gas molecules within pores can move quickly in comparison to the time it takes for heat to be conductively transported through the solid. Mass and energy each can be described with advective and diffusive transport equations, but the diffusion coefficients and Peclet numbers (describing the ratio of advection to diffusion rates) differ.
The different timescales present in a multi-phase transport model present a challenge for numerical integration but can also help us simplify the transport equations of the model. 

We use a 1-dimensional model to describe a porous body as a function of depth, $x$, and time, $t$. The surface is at $x=0$, and depth corresponds with increasingly positive values of $x$. Bulk quantities that are a function of depth and time ($x,t$) are temperature $T$, bulk mass density $\rho_\text{eff}$, bulk specific heat capacity $c_{p,\text{eff}}$, bulk thermal conductivity, $k_\text{T,eff}$ and porosity $\phi$. Here, $\rho_\text{eff}$ is the mass per unit volume including all solids and gases. Similarly $c_{p,\text{eff}}$ and $k_\text{T,eff}$ are effective specific heat capacity and thermal conductivity per unit volume including all constituents. 

The solid material consists of a rocky substrate and one or more ices. An integer index `$i$' is used to label solid and gas constituents. Each volatile is associated with an ice. The pores contain molecular gas species, each of which is described with a number density and a partial pressure $P_i$. 
To calculate parameters that rely on the volume physically occupied in each pore, we scale bulk values by the local porosity $\phi$, which is the ratio of pore volume to total volume. Quantities that rely on the volume physically occupied by solid material are scaled by $(1-\phi)$. 
We denote the bulk number density of a gas species `$i$' per unit volume as $n_{g,i}$. 
The number density of that gas species within a pore is then $n_{g,i}/\phi$ giving a relation between partial pressure, temperature and bulk number density that is based on the ideal gas law
\begin{align}
P_i = \frac{n_{g,i} }{\phi} k_B T . 
\label{eqn:Pi}
\end{align}

Additional details of the numerical method are described in \ref{App:Num_model}.

\subsection{Gas transport in the Knudsen Regime}
\label{subsubsec:Knudsen_Regime}


In a porous medium described by a mean pore diameter, $d_p$, the Knudsen number is the dimensionless ratio of the mean free path $\lambda$ in the gas to the pore diameter 
\begin{equation}
 \text{Kn} \equiv \frac{\lambda}{d_p}. \label{eq:Knudsen}
\end{equation}
We estimate the mean free path in the gas 
\begin{equation}
\lambda \sim \frac{1}{\sqrt{2} n_g \sigma} = \frac{k_BT}{P\sqrt{2}\sigma},
\label{eq:mfp_eq}
\end{equation}
where $n_g$ is the number density of molecules in the gas, and on the right, we have used the ideal gas law. Here, $k_B$ is Boltzmann's constant.
The factor of $\sqrt{2}$ comes from kinetic theory \citep{Chapman_1990}. 
Using a collisional cross section typical of collisions between two CO$_2$ molecules with $\sigma_{\text{CO}_2}\sim 2\times10^{-19}\ \text{m}^2$ \citep{Song_2024}, similar to that of molecular hydrogen, we estimate a mean free path of %
\begin{equation}
\lambda \sim 0.15 \ \text{m} \left(\frac{T}{300 \ \text{K}}\right)\left(\frac{1 \text{ Pa}}{P}\right)\label{eq:mfp_est}.
\end{equation} 
We have used a pressure similar to that estimated in the midplane at 1~AU from the Sun for a proto-solar disk (following Table 1 by \citealt{Quillen_2024}). 

There are few constraints on the pore distribution of primordial porous bodies; consequently, we follow \citet{Bouziani_2022} who described the porous medium with a mean pore diameter $d_p$ and considered a large range of possible values, with $d_p$ in the range $10^{-2}$ to $10^{-8}$ m. 

As long as the pressure remains low, Equation \ref{eq:mfp_est} implies that the mean free path should exceed the pore diameter, $\lambda \gg d_p$, giving Knudsen number $\text{Kn} \gg 1$. 
In this regime, called the Knudsen regime, molecule-wall collisions should be more frequent than collisions between gas molecules.
Consequently, the flux of each molecular species through the porous medium can be described independently of the other species and with a diffusion coefficient, known as the Knudsen diffusion coefficient, that depends upon the thermal velocity and the pore diameter \citep{Welty_2008}. 
Knudsen diffusion has also used in comet out-gassing models \citep{Fanale_1984,Bouziani_2022}.



For a molecule of type indexed `$i$', and with 
number of molecules per unit volume $n_{g,i}$,
the equation for gas transport in the Knudsen regime is:
\begin{equation}
 \frac{\partial {n}_{g,i}}{\partial t} = \frac{\partial}{\partial x}\left(D_{\text{Kn},i}\frac{\partial n_{g,i}}{\partial x}\right) + \dot{n}_{\text{sub},i}.
\label{eq:mass_transport}
\end{equation}
The rate $\dot n_{\text{sub},i}$ is the rate that 
the number density is increased by local sublimation of the associated ice (the solid form of the molecule). If the species is deposited on surfaces as ice, then $\dot n_{\text{sub},i}$ would be negative. We use a subscript `$g$' to make it clear that $n_{g,i}$ refers to the number density of a gas molecule rather than that of a solid component. 
The Knudsen diffusion coefficient $D_{\text{Kn},i}$ depends on the product of the pore diameter and the thermal velocity 
\citep{Welty_2008}; 
\begin{equation}
 D_{\text{Kn},i} = \frac{d_p V_{\text{th},i}(T)}{3}.
 \label{eq:Knudsen_Diffusion}
\end{equation}
The mean of the magnitude of the velocity (computed using the Boltzmann distribution) or thermal velocity $V_{\text{th},i}$ is 
\begin{equation}
 V_{\text{th},i}(T) = \sqrt{\frac{8 k_B T}{\pi m_i}},
 \label{eq:V_th}
\end{equation}
where $m_i$ is the mass of a single molecule of type `$i$'. 

\subsection{Energy Transport}
\label{sec:Energy}



To model energy transport, we adopt a multiphase multi-constituent model for transport through a porous medium similar to that by \cite{Warning_2015};
\begin{align}
 \rho_\text{eff}c_{p,\text{eff}} \frac{\partial T}{\partial t} & = \frac{\partial}{\partial x}\left(k_\text{T,eff}\frac{\partial T}{\partial x}\right) 
 - \sum_{i} c_{p,g,i }J_i \frac{\partial T}{\partial x} \nonumber \\
 & \ \ \ \ - \sum_i E_{\text{sub},i}\dot{n}_{\text{sub},i}.
\label{eq:Energy}
\end{align}
The term on the left is the energy required per unit volume to increase the local temperature. 
The term proportional to $k_{T,\text{eff}}$ describes diffusion of energy due to thermal conductivity. 
The rightmost term, $\sum_i E_{\text{sub},i}\dot{n}_{\text{sub},i}$, is the energy per unit volume per unit time from the sublimation and deposition of volatiles. The quantity $E_{\text{sub},i}$ is the latent heat per molecule of sublimation of the volatile species labeled by subscript $i$. The bulk number density rate $\dot n_{\text{sub},i}$ is the number of molecules of type `$i$' per unit time per unit volume that are sublimating or being deposited in the porous medium. 

The second term from the right in Equation \ref{eq:Energy}, (proportional to $\frac{\partial T}{\partial x}$), is advective and depends on the fluxes $J_i$, of each type of molecule through the porous medium. The quantities $c_{p,g,i}$ are the specific heat capacity of each gaseous molecular component. We will discuss the advective term further in Section \ref{subsubsec:Advective_heat}. 
Neglecting the rightmost two terms, the energy transport equation of Equation \ref{eq:Energy} describes diffusive thermal heat transport in a solid. 

The bulk density, bulk specific heat capacity,
and bulk thermal conductivity ($\rho_\text{eff}, c_{p,\text{eff}}$, $k_{T,\text{eff}}$) take into account both solid and gas constituents in a small local volume. 
The bulk density times effective specific heat capacity $\rho_\text{eff}, c_{p,\text{eff}}$ is related to solid and gas constituents by 
\begin{align}
 \rho_\text{eff}c_{p,\text{eff}} &= (1-\phi)\sum_{i,\text{solids}}(\chi_{s,i}\rho_{s,i} c_{p,s,i}) 
 \nonumber \\
 & \ \ \ + \phi\sum_{i,\text{gases} }(\chi_{g,i}\rho_{g,i} c_{p,g,i}) .\label{eqn:rhocp}
\end{align}
Here $\chi_{g,i}$ denotes the gas species volume fraction in pores, where the subscript `$i$' indicates the species, and the subscript `$g$' indicates gas. The density $\rho_{g,i}$ is the gas mass density of species $i$ inside the pores. Similarly, $c_{p,g,i}$ is its specific heat capacity. 

The quantity $\chi_{s,i}$ is the volume fraction of a solid species `$i$' in the matrix excluding the pores. Again the subscript $i$ labels the species and the subscript `$s$' indicates that it is a solid. 
The specific heat capacity $c_{p,s,i}$ is that of the particular solid constituent, 
and $\rho_{s,i}$ is its density.
These quantities are computed for the matrix material that excludes the pores. 
For each volatile molecule we can use the same index to represent quantities for both gas and solid ice phases of the molecule. 

We estimate the bulk thermal conductivity as 
\begin{align}
 k_\text{T,eff} &= (1-\phi)\sum_{i,\text{solids}}(\chi_{s,i}k_{T,i}) ,\label{eqn:k_eff} 
\end{align}
where $k_{T,i}$ is a solid constituent's thermal conductivity. 
We neglect the contribution of radiation inside pores. Transport of energy by the gas is described by the advective term in Equation \ref{eq:Energy}. 


\subsubsection{Volumetric Heat Capacity}
\label{subsubsec:volumetric_heat}

In this subsection we show that 
Equation \ref{eqn:rhocp} can be simplified to only account for the solid material. We adopt the near-universal value for the volumetric heat capacity of solids (the product of a material's density $\rho$ times its specific heat capacity $c_p$), 
\begin{equation}
C_\text{uni} = \rho c_p \approx 3 \times 10^6 \ \text{J m}^{-3} \text{ K}^{-1}.
\label{eq:rhocp_uni}
\end{equation} 
\citep{Materials_book}. 
An ideal gas has heat capacity per molecule at constant pressure 
\begin{equation}
C_p = \frac{\gamma}{\gamma - 1} k_B, 
\end{equation}
where $\gamma$ is the adiabatic index. 

We compare the volumetric specific heat of the gas of species $i$ in a pore $c_{p,g,i}\rho_{g,i}$ with the universal value $C_\text{uni}$ for a the solid. 
With adiabatic index $\gamma_i$ for the species, the volumetric specific heat of gas in a pore is 
\begin{align}
\rho_{g,i} c_{p,g,i} & \sim \frac{P_i}{T} \frac{\gamma_i}{\gamma_i -1 }\nonumber \\
 &= 10^{-3} \text{ J} \text{ m}^{-3} \text{ K}^{-1} \! \frac{\gamma_i}{(\gamma_i - 1)} \left(\frac{P_{i}} {1 \ \text{Pa}}\right) \nonumber \\& \ \ \ \times \left(\frac{T}{1000\ \text{K}}\right)^{\! -1}\!\!\!.
 \label{eq:C_pn_i} 
\end{align}
The gas pressure in a circumstellar disk is usually below a few Pa (e.g., see Table 1 by \citealt{Quillen_2024} for a protosolar disk model). 
Consequently, the volumetric heat capacity of the gas 
\begin{align}
\rho_{g,i} c_{p,g,i} \ll C_\text{uni}
\end{align} is likely to be many orders of magnitude below the universal value for a solid given in Equation \ref{eq:rhocp_uni}. The energy required to heat or cool the gas in the pores is likely to be negligible when compared to that required to heat or cool the solid components. We can neglect the gas contribution to $\rho_\text{eff} c_{p,\text{eff}}$. Specifically, we neglect the rightmost term in Equation \ref{eqn:rhocp}. 



\subsubsection{Advective Heat Transport}
\label{subsubsec:Advective_heat}

The energy transport equation of Equation \ref{eq:Energy} contains an advective heating term proportional to $\frac{\partial T}{\partial x}$ due to gas flowing through the porous medium. We assess the strength of the advective heating term in comparison to the diffusive term in Equation \ref{eq:Energy} by computing the dimensionless ratio known as the P\'eclet number. 

The P\'eclet number for heat transfer is the ratio of the approximate sizes of the diffusive and advective terms in a heat equation. 
In the Knudsen regime, 
the flux of molecular constituent `$i$' through the porous medium is 
\begin{align}
J_i \sim D_{\text{Kn},i} m_i \nabla n_{g,i}.
\end{align}
We approximate the gradient with a length scale $L$ giving $\nabla n_{g,i} \sim n_{g,i}/L$. 
The gradient in the strength of the diffusion term in Equation \ref{eq:Energy} is also approximated by the length $L$. 
For the energy transport equation (Equation \ref{eq:Energy}), the P\'eclet number is 
\begin{align}
 \text{Pe} \approx\frac{1}{k_{T,\text{eff}}} \sum_{i} c_{p,g,i} \rho_{g,i} D_{\text{Kn},i} . 
 \label{eqn:Pe}
\end{align}

To estimate the P\'eclet number, we use an approximate value for the thermal conductivity of a rocky solid that is similar to that of glass or concrete, 
\begin{align}
k_\text{T,typ} = 1 \text{ W m }^{-1} \text{ K}^{-1} \label{eqn:kuni}
\end{align}
\citep{Bird_2007}. 
This value is about half the conductivity value of water ice \citep{Bonales_2017}.
We use Equation \ref{eq:Knudsen_Diffusion} for the Knudsen diffusion coefficient which depends on the thermal velocity in Equation \ref{eq:V_th}. To give a high estimate for the diffusion coefficient, we use the mass of molecular hydrogen and temperature exceeding the peak temperature in any of our simulations, $T =$ 1000 K.
We use Equation \ref{eq:C_pn_i} to estimate 
$\rho_{g,i} c_{p,g,i}$ for molecular hydrogen. 
Putting these values into Equation \ref{eqn:Pe}, the resulting estimate for the P\'eclet number is 
\begin{align}
 \text{Pe} 
 & \sim 10^{-6} \left( \frac{P }{1 \text{ Pa}} \right)
 \left(\frac{T}{1000 \text{ K}} \right)^{\frac{3}{2}}
\left( \frac{d_p}{10^{-6} \text{ m}} \right) \nonumber \\
& \ \ \ \ \ \times 
\left(\frac{m}{2 \text{ AMU}} \right)^{-\frac{1}{2}}. 
\label{eqn:Pec}
\end{align}
Because this is based on molecular hydrogen we have dropped the $i$ indices, pressure $P$ represents the total gas pressure, $m$ is the mass of molecular hydrogen, and AMU refers to an atomic mass unit. 

Equation \ref{eqn:Pec} shows that the P\'eclet number is likely to be small. This implies that we can neglect the advective term (proportional to $\frac{\partial T}{\partial x}$) in the energy transport equation (Equation \ref{eq:Energy}).

\subsubsection{Simplified Heat Equation}

In Section \ref{subsubsec:volumetric_heat}, we showed why only the heat capacity of solid materials needs to be taken into account. In Section \ref{subsubsec:Advective_heat}, we found that we can neglect the advective heating term in Equation \ref{eq:Energy} because the P\'eclet number is small. We use the near universal value $C_\text{uni}$ to describe the specific heats of all the solid constituents. 
With these simplifications, we reduce the energy transport equation (Equation \ref{eq:Energy}) to 
\begin{equation}
 (1-\phi)C_\text{uni} \frac{\partial T}{\partial t} = \frac{\partial}{\partial x}\left( k_\text{T,eff} \frac{\partial T}{\partial x}\right) - \sum_i E_\text{sub}\dot{n}_{\text{sub},i}.
\label{eq:final_heat}
\end{equation}
We have scaled $C_\text{uni}$ to take into account the fraction of volume $(1- \phi)$ occupied by solid material. 



\subsubsection{Sublimation, Deposition, \& Vapor Pressure}

We use a non-equilibrium relation to estimate the sublimation/deposition rate of a vapor. 

The Hertz-Knudsen adsorption rate describes the rate of adsorption of gas molecules on a surface. For a gas with pressure $P$, the Hertz-Knudsen rate of adsorption per unit area is 
\begin{align}
\frac{dN}{dA\ dt} \sim \frac{\alpha_{\text{ads}} P}{\sqrt{2\pi m k_B T}} ,\label{eqn:dA0}
\end{align}
where $m$ the mean molecular mass. The dimensionless coefficient $\alpha_\text{ads}$ can be used to fit experimental data or theoretical predictions and depends on properties of the molecule and (e.g., \citealt{Kossacki_1999}). 

A non-equilibrium rate for sublimation or deposition on a surface per unit area can be estimated by modifying the Hertz-Knudsen adsorption rate so that 
the rate of sublimation or deposition on a surface per unit area depends on 
the difference between the partial and vapor pressure (e.g., \citealt{Scarpa_2002,Warning_2015,Kossacki_2021}). 
For a molecular species with index `$i$', the modified adsorption rate is
\begin{align}
\frac{dN_i}{dA\ dt} \sim \frac{\alpha_{\text{ads},i} (P_i - P_{v,i} )}{\sqrt{2\pi m_i k_B T}} ,\label{eqn:dA}
\end{align}
which is based on the 
number $N_i$ of molecules of type `$i$' impacting the surface per unit of time \citep{Holyst_2015}. 


We estimate that the number of pores per unit volume scales with the pore size, $N_\text{p} \propto d_p^{-3}$ with porosity $\phi \approx N_\text{p} d_p^3$. 
Following \citet{Schweighart_2021}, we approximate the surface area in a volume element with the product of the number of pores times their surface area $\sim N_\text{p} d_p^2 \sim \phi/d_p$. 
The surface area in a volume element and Equation \ref{eqn:dA} give an adsorption or deposition rate per unit volume of 
\begin{align}
\dot n_{\text{dep},i} &\sim \frac{\phi (P_{i} - P_{v,i})}{\sqrt{2\pi A_i m_i k_B T} } 
\frac{\alpha_{\text{ads},i}}{ d_p} \nonumber \\
&\sim V_{\text{th},i} (n_{g,i}-n_{v,i})\frac{\alpha_{\text{ads},i}}{4d_p},
\label{eqn:subrate}
\end{align}
where we have used Equation \ref{eq:V_th} for the thermal velocity and defined a bulk number density associated with the vapor pressure 
\begin{align}
n_{v,i} \equiv \frac{\phi P_{v,i}}{k_B T}.
\end{align}
If the partial pressure exceeds the vapor pressure and Equation \ref{eqn:subrate} gives a positive number, then we assume that ice is deposited. If Equation \ref{eqn:subrate} is negative then we assume that ice is sublimated and we estimate the sublimation rate as 
\begin{align}
\dot n_{\text{sub},i} =- \dot n_{\text{dep},i}.
\end{align}

We adopt the same constant value of $\alpha_{\text{ads},i}$ for all molecular species and neglect the dependence of these coefficients on temperature \citep{Kossacki_1999,Kossacki_2017}.


\begin{table*}[t]
\centering
\caption{Properties of Each Molecular Species}
\begin{tabular}{lllllll}
\hline
Quantity & Symbol & Units & H$_2$O & CO & CO$_2$ & Forsterite \\
\hline
Mean Molecular Weight & $A_i$ & atomic mass & 18 & 28 & 44 & 140.691\\ 
Density of Associated Solid & $\rho_{s,i}$ & kg m$^{-3}$ & $10^3$ & 870 & 1640 & 3270\\
Latent Heat of Sublimation & $H_{\text{sub},i}$ & J kg$^{-1}$ & 28.4$\times 10^5 $ & 2.16 $\times 10^5$ 
& 7.58 $\times 10^5$ & -- \\
Sublimation Energy & E$_{\text{sub},i}$ & J/particle& $8.489 \times10^{-20}$ & $1.004\times 10^{-20}$ & $5.538 \times10^{-20}$ & --\\
Adsorption Coefficient & $\alpha_\text{ads}$ & -- & 0.1 & 0.1 & 0.1 & -- \\
Solid Thermal Conductivity & $k_{T,i}$ & W m$^{-1}$ K$^{-1}$ & 2.1 & 0.7 & 0.7 & 5 \\
Coefficient for Vapor Pressure & $A_{v,i}$ &- & 25.279 & 21.7629 &30.123 & -- \\
Coefficient for Vapor Pressure & $B_{v,i}$ &K & 5132.0 & 748.2 &4154.0 & --\\
\hline 
\end{tabular}
\caption*{\newline 
The CO ice density is by \citet{Gonzalez_2022}, CO$_2$ ice density is by \citet{Leliwa_2013}, and Forsterite solid density is by \citet{anthony2001handbook}.
The latent heat of sublimation for water ice is consistent with the value of $2.59 \times 10^6$ J/kg by \citet{Leliwa_2013} and the latent heats of sublimation for CO and CO$_2$ are by \citet{Leliwa_2013}. \citealt{Kossacki_2014} show temperature dependence for the adsorption coefficient for H$_2$O, finding a minimum value near $\alpha_\text{ads} = 0.1$ (in equation \ref{eqn:dA0}) when the temperature exceeds the sublimation temperature. We take this value for all volatiles tested when sublimation or deposition is occurring. The CO solid thermal conductivity is by \citet{Romanova_2014}, CO$_2$ solid thermal conductivity is by \citet{Saiduzzaman_2025}, and Forsterite thermal conductivity is by \citet{Clauser_1995}. The coefficients $A_{v,i}$ and $B_{v,i}$ for the vapor pressure curve are defined in Equation \ref{eqn:Pvk2} and are based on fits to laboratory measurements compiled by \citet{Fray_2009}.
}
\label{tab:MolecularComponents}
\end{table*}

For each volatile molecular species, we adopt a model for vapor pressure consistent with the Clausius-Clapeyron relation. We assume vapor pressure as a function of temperature for species with index `$i$' can be described by the function 
\begin{align}
P_{v,i}(T) = e^{A_{v,i} - B_{v,i}/T}. \label{eqn:Pvk2}
\end{align}
Table \ref{tab:MolecularComponents} lists
coefficients $A_{v,i}$ and $B_{v,i}$ for water, carbon dioxide and carbon monoxide based on fits to laboratory measurements by \citet{Fray_2009}. 

\subsection{Porosity}
\label{subsec:porosity}
Porosity is a measure of the percentage of void space within a given volume. In our simulations, because ices are sublimated or deposited, the porosity varies as a function of depth and time. 
We describe our methods for computing the porosity as a function of depth in \ref{App:P_Phi_Solids}.
While we allow the porosity to vary, we hold the pore diameter constant.

\subsection{Boundary Conditions}

At the base of the array (the grid point furthest from the surface), we adopt a zero heat flux or Neumann boundary condition.
Similarly, we adopt a zero diffusive flux lower boundary condition for each gas species. 


Comet models (e.g., \citealt{Gombosi_1986}) adjust the surface temperature so that 
the heat flux from absorbed solar radiation is equal to the sum of the conductive heat flux
and that emitted via thermal radiation.
Instead of absorbed solar radiation, the surface of our simulated porous body is heated due to thermal exchange by collisions with the surrounding nebular gas, which is dominant over heating due to gas drag \citep{Gombosi_1986,Desch_2002}. 
Following \citet{Desch_2002, Gombosi_1986},
our surface boundary condition for a porous body exposed to the nebula is
\begin{align}
K_{T,\text{eff}} \frac{dT}{dx} \Bigg|_{x=0} & = 
n_\text{neb} V_{\text{th},\text{neb}} k_B (T_\text{neb} - T_\text{surf}) \nonumber \\ & \ \ + \epsilon_\text{surf} \sigma_{SB} (T_\text{neb}^4 - T_\text{surf}^4) 
 \label{eqn:E_boundary}
\end{align}
where $T_\text{surf}$ is the surface temperature and $\epsilon_\text{surf}$ is the emissivity at the surface. 
Quantities describing the nebular gas are number density of molecules ($n_\text{neb}$), temperature ($T_\text{neb}$), and thermal velocity ($V_{\text{th},\text{neb}}$). 

We solved Equation \ref{eqn:E_boundary} numerically for a nebular pressure of a few Pa, and 
we found that the surface temperature remains very close to that of the nebula; $T_\text{surf} \sim T_\text{neb}$. 
Consequently, we 
set both the surface temperature and gas pressure to that of the nebula giving time-dependent and Dirichlet 
surface boundary conditions in both quantities. 

\subsection{Ambient Environmental Conditions in the protoplanetary Nebula}

To model a transient heating event in a protoplanetary disk, we require an ambient or equilibrium state for the protoplanetary disk prior to the transient heating event. 
For the ambient gas disk, we adopt the protosolar disk model by \citet{Quillen_2024}. This model is similar to the protosolar disk model by \citet{Desch_2007} that takes into account subsequent planet formation and migration, but it has power-law exponents describing its dependence on radius that are similar to the minimum mass solar nebula by \citet{Hayashi_1981}. 



This disk model 
has an equilibrium midplane gas density, $\rho_\text{gas}$, gas temperature, $T_\text{gas}$, and gas pressure, $P_\text{gas}$:
\begin{align}
 \rho_{\text{gas}} &= 1.6\times 10^{-8} {\text{ kg m}}^{-3} \left( \frac{a_o}{1 \text{ AU}} \right)^{-\frac{11}{4}} 
 \label{eq:AmbientDensity} \\ 
 T_\text{gas} &= 150 \text{ K} \left( \frac{a_o}{1 \text{ AU}} \right)^{-\frac{1}{2}} 
 \label{eq:AmbientTemperature} \\
 P_\text{gas} &= 9.92 \text{ Pa} \left(\frac{a_o}{1 \text{ AU}}\right)^{-\frac{13}{4}}\left(\frac{\bar{m}}{2 \text{ AMU}}\right)^{-1}
 \label{eq:AmbientPressure} 
\end{align}
where ${\bar{m}}$ is the average mass of particles in the nebular gas and $a_0$ is distance from the central star.

We use these values to determine the number density of the ambient nebular gas. With $X_i$ equal to the abundance of molecular species $i$ with respect to molecular hydrogen, the ambient nebular number density of that same species is
\begin{align}
 n_{\text{neb},i}& =4.8\times10^{18}\text{ particles m}^{-3} \left(\frac{a_o}{1 \text{ AU}}\right)^{\!-\frac{11}{4}}\nonumber \\
 & \ \ \ \times \left(\frac{\bar{m}}{2 \text{ AMU}}\right)^{-1}X_i.
\end{align}


\subsubsection{Transiently Heating Nebular Gas and Decay Rate}

We model a transient heating event in the nebula using an instantaneous increase in nebular temperature and density, at $t=0$, followed by an exponential decay back to the initial conditions. This is an 
approximation to the conditions experienced by an object embedded in a disk undergoing 
a nebular shock (e.g., \citealt{Ciesla_2002, Desch_2002}, also see \citealt{Miura_2005}). 

We model the transient heating event with
\begin{equation}
 X(t) = X_\text{ambient} + (X_\text{max} - X_\text{ambient})e^{-t/\tau}
\label{eq:DecayRate}
\end{equation}
where $X$ is a representative variable for any decaying nebular environmental quantity ($T_\text{neb}$, $\rho_\text{neb}$, X$_{volatile}$) and $\tau$ is the decay timescale. We calculate the nebular gas pressure during this decay period using the ideal gas law. 

We set the ambient volatile gas fraction for the gas species being simulated to be 0, with the maximum gas fraction being listed in Table \ref{tab:Runs}. We set the $H_2$ gas fraction as the rest of the available gas fraction (X$_{H_2}$ = 1-X$_{volatile}$).

The strength of an event is described with the ratio of the chosen maximum temperature $T_\text{neb,max}$ to the ambient temperature $T_\text{neb,ambient}$:
\begin{align}
\Gamma \equiv \frac{T_\text{neb,max}}{T_\text{neb,ambient}}.
\end{align}
Ambient nebular values are calculated using Equations \ref{eq:AmbientDensity}, \ref{eq:AmbientTemperature}, and \ref{eq:AmbientPressure}. Ambient nebular values at our fiducial orbital radius of 5 AU can be found in Table \ref{tab:Fiducial}. 



\section{Numerical Results}
\label{sec:Results}

We simulate a cool, porous body exposed to a transiently heated nebular environment. We increase the ambient values of gas temperature and number density by a factor of the event's strength, 
$\Gamma$, and increase the pressure by $\Gamma^2$. We assume a decay timescale, $\tau$, (Seen in Equation \ref{eq:DecayRate}) to allow the environmental values of gas temperature and number density to decay back to their ambient values, calculating pressure with the ideal gas law. 

For these simulations, we fix the pore diameter. As deposited ice can decrease pore diameter, this is only a good approximation if the ice fraction is initially low. 

The depth required for a simulation is dependent on the thermal skin depth of the body which depends on the duration $t_{heat}$ of the heating event. To order of magnitude, the skin depth is 
\begin{align}
l_\text{skin}(t) & \sim \sqrt{\frac{k_{T}t_{heat}}{C_\text{uni} }} \nonumber \\
&\sim 2.5 \text{ cm} \left( \frac{t_{heat} }{ 360 \text{ sec} } \right)^\frac{1}{2},
\label{eq:skin_depth}
\end{align}
and we have used a heat diffusivity $D_\text{heat} = k_{T,\text{eff}}/C_\text{uni}$ with volumetric heat capacity $C_\text{uni}$ given in Equation \ref{eq:rhocp_uni} and the thermal conductivity $k_{T}$ of Forsterite, listed in Table \ref{tab:MolecularComponents}.
For a 6 minute heating event, the skin depth is about 2.5 cm.
In this paper, we present short simulations (360 sec) with the goal of understanding phenomena associated with transient heating. 
After the heated medium returns to its cooler ambient level, the deposition front can continue to advance, eventually reaching a final position that is deeper than the skin depth. To see the final resting place of the deposition front, we choose a grid length that is a few times larger than the skin depth. 
The integration duration is chosen so that in most cases, the simulation is integrated long enough that simulated deposition fronts reach their final position, with exceptions for the largest tested $\Gamma$ and the longest duration $\tau$, for which the movement of the ice deposition front has decreased but not ceased.

Fiducial parameters for our simulations are listed in Table \ref{tab:Fiducial} and parameters specific to individual simulations are in listed in Table \ref{tab:Runs}.
Each simulation tracks a single volatile molecular species in both solid and gas forms, molecular hydrogen, and a rocky substrate. The properties of the volatiles are listed in Table \ref{tab:MolecularComponents}, and those of the rocky substrate are in the forsterite column in Table \ref{tab:MolecularComponents}. 

We plot snapshots of the deposition front for individual runs. We also run a suite of simulations varying individual parameters, where we plot the depth of the ice front and the mass of ice deposited per square meter of surface area. 

We explore how the deposition front is sensitive to the transient heating event's strength ($\Gamma$), pore diameter ($d_p$), ice volume fraction ($f_\text{ice}$), decay timescale ($\tau$), porosity ($\phi$), and the particular volatile species that is included in the simulation.

\begin{table*}[t]
 \centering
 \caption{Fiducial Values}
 \begin{tabular}{llll}
 \hline Parameter & Variable & Value & Reference \\\hline
 Orbital Radius & $a_o$ & 5 [AU] & \\
 Initial Nebula Gas Density & $\rho_{gas}$($a_o$=5 AU) & $\sim 1.9\times10^{-10}$ [kg m$^{-3}$]& Equation \ref{eq:AmbientDensity}\\
 Initial Temperature & $T_{gas}$($a_o$=5 AU) & $\sim 67$ [K] & Equation \ref{eq:AmbientTemperature}\\
 Initial Porosity & $\phi$ & 0.5 & \citet{Ballouz_2026,Ryan_2026}\\
 Pore Diameter & $d_p$ & $10^{-6}$ [m] & \\
 Transient Heating Event's Strength & $\Gamma$ & 5 & \\
 Integration Duration & &360 [sec]& \\
 Integration Time-step & $\Delta t$ & & \ref{App:Num_model}\\ 
 Decay Timescale & $\tau$ & 360 [sec] & \\ 
 Grid Length & & 0.04 [m]& \\ 
 Grid Resolution & $\Delta x$ & 0.001 [m] & \\ 
 \hline
 \end{tabular}
 \label{tab:Fiducial}
\caption*{\newline 
We set the initial gas pressure within the body to be the vapor pressure of the volatile gas (Equation \ref{eqn:Pvk2}). After the simulation starts, there is a short transient period where the pressure, gas number density, and temperature return to ideal gas conditions, which can be seen as the jump in pressure at t = 0 sec and t = 30 sec in Figures \ref{fig:Dry_Body}, \ref{fig:Icy_Body}, \ref{fig:Icy_Pore_Radius}. We set the initial porosity to $\phi = 0.5$, agreeing with the low-end estimates of asteroid porosities from \citet{Ballouz_2026, Ryan_2026}.}
\end{table*}

\begin{table*}[]
 \centering
 \caption{Simulation Runs With Initial Parameters}
 \begin{tabular}{c||cccccccccc}
 \hline Run & Increment & $a_o$ [AU] & $d_p$ [m] & $\phi$ & $\Gamma$ & $\tau$ [s] & $f_\text{ice}$ & Volatile & Gas& Figure \\ 
 Number &&&&&&&&Species&Fraction \\\hline
 1 & - & 5 & $10^{-6}$ & 0.5 & 5 & 360 & 0.00 & H$_2$O & $10^{-4}$ & \ref{fig:Dry_Body}\\ 
 2 & - & 5 & $10^{-6}$ & 0.5 & 5 & 360 & 0.01 & H$_2$O & $10^{-4}$ & \ref{fig:Icy_Body},\ref{fig:Icy_Pore_Radius},\ref{fig:vary_ice_type} \\ 
 3 & - & 5 & $10^{-8}$ & 0.5 & 5 & 360 & 0.01 & H$_2$O & $10^{-4}$ & \ref{fig:Icy_Pore_Radius}\\ 
 4 & - & 5 & $10^{-6}$ & 0.5 & 10 & 360 & 0.01 & H$_2$O & $10^{-4}$ & \ref{fig:Ice_Frac_Shock_Strength}\\ 
 5 & - & 5 & $10^{-6}$ & 0.5 & 5 & 360 & 0.01 & CO$_2$ & $10^{-5}$ & \ref{fig:vary_ice_type} \\ 
 6 & - & 55 & $10^{-6}$ & 0.5 & 5 & 360 & 0.01 & CO & $10^{-5}$ & \ref{fig:vary_ice_type}\\ 
 \hline
 7-11 & $\times 0.1$ & 5 & $10^{-5}-10^{-9}$ & 0.5 & 5 & 360 & 0.01 & H$_2$O & $10^{-4}$ & \ref{fig:vary_pore_radius_H2O_icy} \\ 
 12-20 & 1 & 5 & $10^{-6}$ & 0.5 & 2-10 & 360 & 0.01 & H$_2$O & $10^{-4}$ & \ref{fig:vary_shock_H2O_Icy}\\ 
 21-26 & 360 & 5 & $10^{-6}$ & 0.5 & 5 & 360-2160 & 0.01 & H$_2$O & $10^{-4}$ & \ref{fig:vary_tau}\\ 
 27-35 & 0.05 & 5 & $10^{-6}$ & 0.9-0.5 & 5 & 360 & 0.01 & H$_2$O & $10^{-4}$ & \ref{fig:dual_porosity},\ref{fig:vary_porosity}\\\hline
 \end{tabular} \\ 
 \caption*{\newline The parameters listed are simulation orbital radius, pore diameter ($d_p$), transient heating event's strength as a multiplier of the temperature ($\Gamma$), Decay Timescale ($\tau$), initial ice volume fraction of the body ($f_\text{ice}$), the volatile species being simulated, and the fraction of the environmental gas for which the volatile comprises. For Runs 7 - 11, we set the pore diameter, $d_p$, to different values, in increments of an order of magnitude. 
 The gas fractions used are motivated by \citealt{Du_2014} (H$_2$O) and \citealt{Bosman_2018} (CO, CO$_2$).
Runs 3 and 4 are identical to Runs 10 and 20, however we list them separately for ease of viewing due to being frequently referred to.}
 \label{tab:Runs}
\end{table*}

\begin{figure}[!t]
 \centering
 \includegraphics[width=9cm]{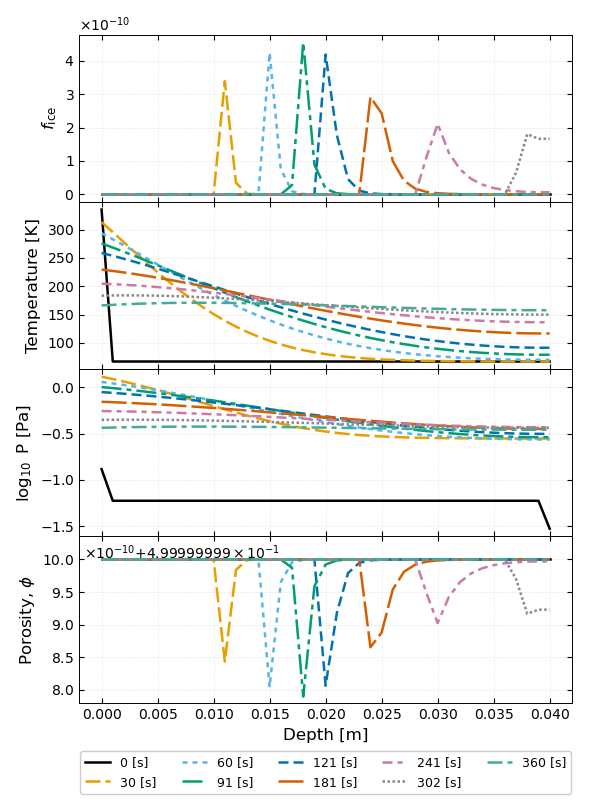}
 \caption{This plot shows the results of Run 1 (with parameters listed in Table \ref{tab:Runs}), an initially dry body is exposed to a heated environment containing water vapor. 
 The x-axis is depth from the body's surface. From top to bottom, panels show ice fraction, gas temperature, $\log_{10}$ pressure within the pores, and porosity. The plotted pressure is the total pressure within the pores, not the partial pressure of the volatile. Each curve shows a different simulation time with 0 sec corresponding to the time that the ambient temperature begins to rise. At t = 0 sec, the initial pressure is set at the vapor pressure of the volatile component. There is a short transient period where the pressure, gas number density, and temperature adjust to the ideal gas law, which is achieved by t = 30 sec. At each time, there is a peak in the ice fraction corresponding to an ice deposition front that advances into the body. As expected, the ice fraction and variation in porosity are inversely proportional, as the amount of rocky material is fixed. 
 }
 \label{fig:Dry_Body}
\end{figure}

\subsection{An Initially Dry Body}

We discuss our numerical simulations of a cool, porous, initially dry body exposed to a suddenly hot protoplanetary nebula that contains molecular hydrogen and warm H$_2$O vapor. Parameters for this simulation, denoted Run 1, are listed in Tables \ref{tab:Fiducial} and \ref{tab:Runs}. 

The panels in Figure \ref{fig:Dry_Body} from top to bottom show the volume fraction of ice ($f_\text{ice}$), the temperature (T), the log of total pressure ($\log_{10}P$), and porosity ($\phi$). All quantities are plotted as a function of depth and at 9 different times, only plotting times during the heating event, $t < \tau$. 

To calculate the ice volume fraction, we first calculate the number density fraction of a solid species (denoted by index $i$) compared to that same solid species assuming there is no void space,
\begin{equation}
 f_{n,i} = \frac{n_{s,i} m_{i}}{\rho_{i}}.
\end{equation}
This is the same as a volume fraction due to the sizes of molecules being compared being equal. Due to being a porous body, we account for void space by dividing this volume fraction by the total volume fraction occupied by solids,
\begin{equation}
 f_\text{ice} = \frac{f_{n,\text{ice}}}{f_{n,\text{ice}}+f_{n,\text{rock}}}.
 \label{eq:f_ice}
\end{equation}

Figure \ref{fig:Dry_Body} shows that as the hot nebular gas enters the cool porous body, an ice deposition front forms. The deposition front is located at a depth where the pressure within the porous body drop below the vapor pressure of water. The peak ice deposition volume fraction is low, never exceeding $10^{-9}$.

When the temperature gradient is large near the deposition front, the temperature at the deposition front is sufficiently below the required temperature for deposition meaning ices are unlikely to diffuse deeper before depositing. The deposition front is initially contained within a narrow spatial band; however, as the temperature gradient decreases, the deposition front widens (see the later times in Figure \ref{fig:Dry_Body} top and bottom panels). When the temperature gradient is low, a wider spatial region is near the sublimation temperature, leading to deposition over a wider range of depth.

With an initially dry porous body the environmental volatiles cause a small increase in the pressure within the porous body, this increase being within an order of magnitude greater than the ambient pressure. The third panel from top in Figure \ref{fig:Dry_Body} shows that the ambient pressure is about 0.1 Pa but pressure in the pores reaches a value that is few Pa.

\subsection{An Initially Icy Body}
\label{subsec:Icy_Body}

\begin{figure}[!t]
 \centering
 \includegraphics[width=9cm]{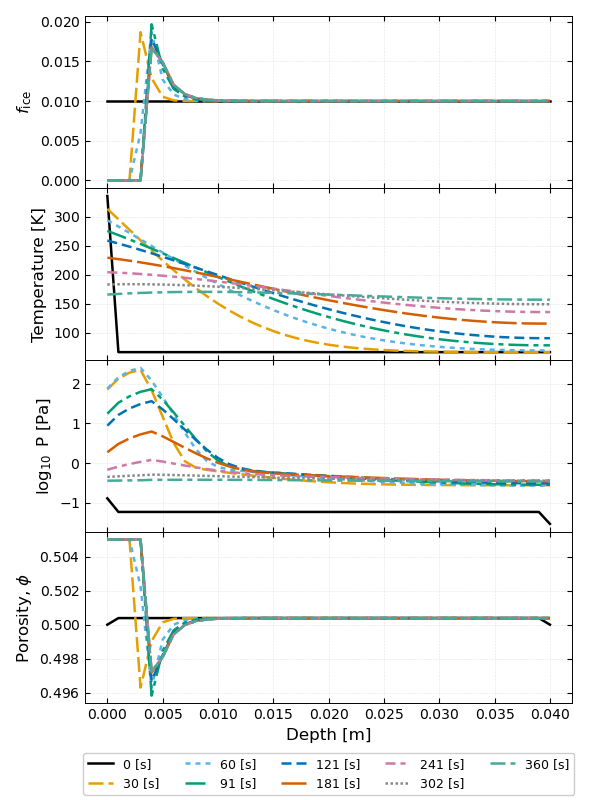}
 \caption{We show a simulation of an initially icy body, denoted Run 2, with parameters listed in in Table \ref{tab:Runs}). This body initially has 1\% ice by volume. This Figure is similar to Figure \ref{fig:Dry_Body} that shows a simulation of an initially dry body. In the top panel, we see an ice deposition layer that is significantly higher in ice volume fraction compared to that seen in Figure \ref{fig:Dry_Body} of the initially dry body. The peak pressure (in the third panel) is much higher in the icy body. The advance speed of the ice deposition layer is slower than in the dry body. In the top 0.5 cm, the medium is dry as ice has been sublimated from the layer. Between a depth of 0.5 cm and 1 cm, sublimated gas has been deposited as ice, increasing the fraction of ice in a thin layer. The sublimation front and ice deposition layer are illustrated in Figure \ref{fig:illust2}. }
 \label{fig:Icy_Body}
\end{figure}

In this section, we discuss a simulation of a cool, porous, and initially icy body exposed to a transiently heated nebular environment. The porous body initially contains 1\% ice by solid volume fraction. The icy body simulation, denoted Run 2, differs from Run 1 only by the initial ice fraction (See Table \ref{tab:Runs} for parameters). The icy body simulation is displayed in Figure \ref{fig:Icy_Body} and is similar to Figure \ref{fig:Dry_Body}.

Figure \ref{fig:Icy_Body} shows that, due to the native ices present within the porous body being incorporated into the deposition front, there is a significant increase in the ice volume fraction ($f_\text{ice}$) at the deposition front. The deposition front advances much slower in an icy body when compared to a dry body. The sublimation of these native ices is associated with a high gas pressure (See Figure \ref{fig:Icy_Body}), with the peak pressure located at the sublimation front exceeding 100 Pa. The peak pressure is much higher than in the simulation of the dry body shown in Figure \ref{fig:Dry_Body}. 

In the icy body simulation we see a sublimation front that advances into the body as the body heats. At depths shallower than the front, the vapor outgasses and the medium is dry because ice that was initially present has been sublimated. At depths deeper than the sublimation front, ice is still present in the body. 
The simulations by \citet{Warning_2015} of microwave heated icy beef under low pressure displayed a sublimation front similar to that seen in our icy body. However the application was freeze drying of meat, so heating took place long enough to entirely dry the substrate. 

At a depth just below the sublimation front our simulation, Figure \ref{fig:Icy_Body} shows a layer where ice has a higher volume fraction than was initially present. Some of the vapor that was sublimated from material near the surface did not escape the body but instead was deposited into this layer. In the dry body simulation, the ice deposition layer consists only of ice deposited from nebular vapor. In the initially icy body, the ice in the deposition layer originates from the body itself. An ice deposition layer was not seen in the comet simulations by \citep{Bouziani_2022}, however their simulations did not allow vapor to penetrate into the porous medium below the sublimation front. 
Simulations of comet out-gassing that allow sublimated material to penetrate below the sublimation front display ice deposition below the sublimation front \citep{Zivithal_2025}. 

Due to out-gassing of sublimated material in confined pore spaces, the pressure (third panel from top) in Figure \ref{fig:Icy_Body} reaches a maximum beyond 100 Pa, exceeding the maximum seen in the dry body. 

\subsubsection{Caveats: Constant Pore Diameter}

In our simulations we adopt a single, fixed pore diameter in a single simulation, but we allow porosity to vary. A more realistic simulation could take into account a distribution of pore sizes and allow variations in mean pore diameter and the pore diameter distribution. We additionally assume all pores are open pores and connected to one another, while a more realistic simulation could allow for some fraction of pores to begin as closed pores, or to allow for pores to become closed due to the ice deposition front.

\begin{figure*}[!ht]
 \centering
 \includegraphics[width=18cm]{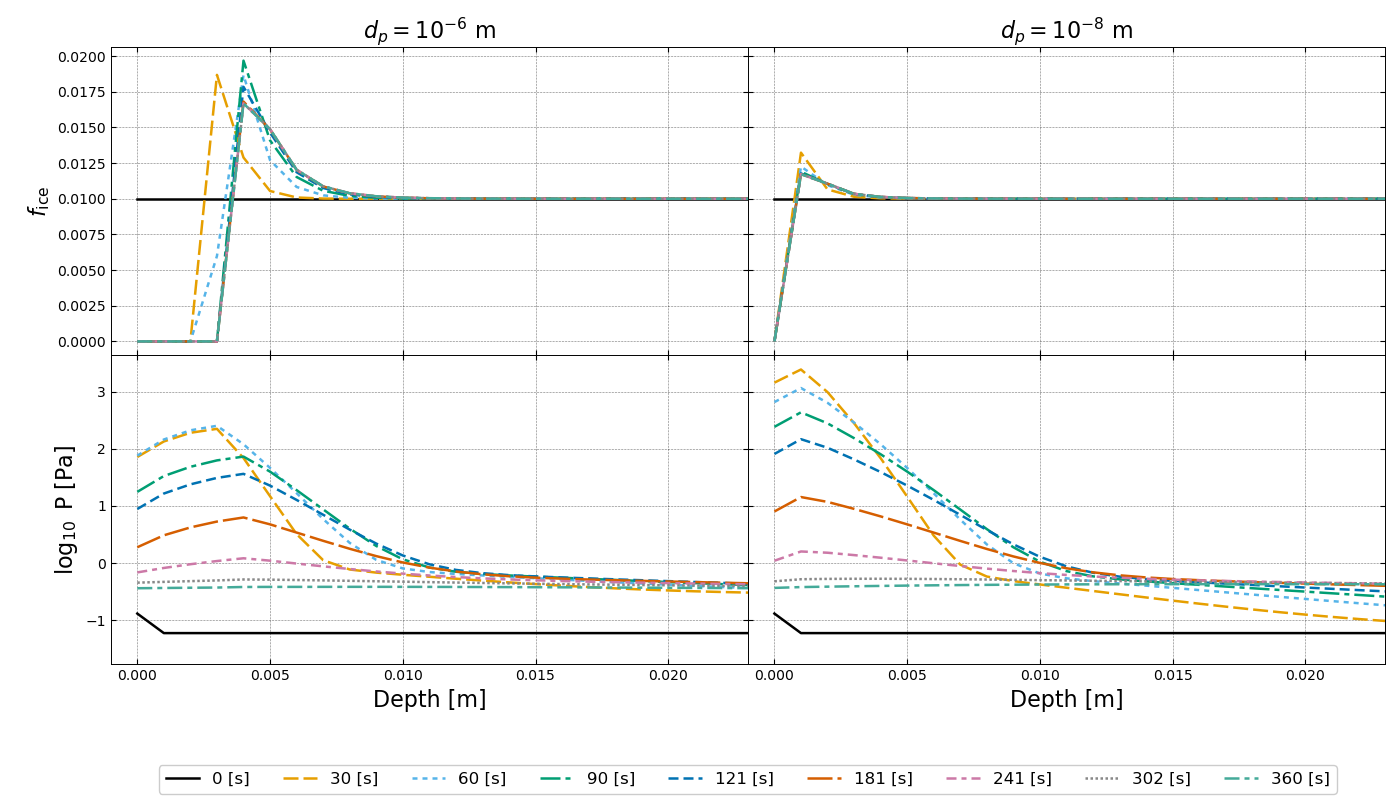}
 \caption{We plot the ice fraction and the internal pressure as a function of distance for Runs 2 and 3, an initially icy body with pore diameters of $d_p$ = $10^{-6}$ m and $d_p$ = $10^{-8}$ m. This icy body is initially 1\% ice by volume. 
 \newline 
 Top Left: We find that as the deposition front moves deeper below the surface over time, the ice contained within the deposition front increases. \newline
 Bottom Left: We find the pressure within the pores increases significantly due to the continuous sublimation of the native ice within the body in contrast with the relatively lower, gradual pressures for an initially dry grain (see Figure \ref{fig:Dry_Body} 
 At any given time, the peak pressure is located near
 the deposition front, indicating that this pressure is due to sublimation of the internal volatiles. \newline
 Top Right: We find that by decreasing the pore diameter, $d_p$, we decrease the depth of the deposition front at any given time when compared to that for a larger pore diameter. With the smaller pore diameters, the deposition front travels more slowly and begins to stop in place.\newline 
 Bottom Right: Decreasing the pore diameter leads to a significant increase in the pressure within the porous body. The peak pressure is located near the peak of the sublimation/deposition front. }
 \label{fig:Icy_Pore_Radius}
\end{figure*}
\subsubsection{Caveats: Solid Ice Fraction}
We limited this study to a low initial solid ice volume fraction of 1\% due to the simplifying assumptions made in Sections \ref{subsubsec:volumetric_heat} -- \ref{subsubsec:Advective_heat}. If the initial ice content is higher, the gas pressure within pores can be high enough that molecule-molecule interactions become important. This regime could be explored in future work.

\subsection{Sensitivity to Pore Diameter}
\label{sec:pore_diameter}

In this section, we run simulations similar to the initially icy body of Run 2, but vary the pore diameter (See Table \ref{tab:Runs}: Runs 2, 3, 7--11). Run 3 is similar to Run 2 but has a pore diameter that is 100 times smaller. Runs 7--11 are similar to Run 2 but have pore diameters spanning $10^{-5}$ to $10^{-9}$ m. 

We simulate two porous bodies with different pore sizes of $d_p = 10^{-6}$ m and $d_p = 10^{-8}$ m (shown in Figure \ref{fig:Icy_Pore_Radius}). With a pore diameter $d_p = 10^{-8}$ m, we find that peak pressure approaches $10^4$ Pa, or 0.1 atm. This is an order of magnitude greater than that of pore diameter $d_p = 10^{-6}$ m. In all cases tested, the peak pressure is near the sublimation front. We find the depth of the deposition front along with the ice fraction peak decreases for smaller pore diameters.

With a pore diameter of $d_p = 10^{-6}$ m, deposition fronts are deeper with a pressure approximately 10 times lower after 30 seconds. While higher pressures can rapidly deposit ice, we find a lower deposition rate at higher pressures, likely attributed to less efficient diffusion through smaller pores. Less diffusion leads to less internal movement of volatile gases after sublimation occurs. 

\begin{figure}[!t]
 \centering
 \includegraphics[width=9cm]{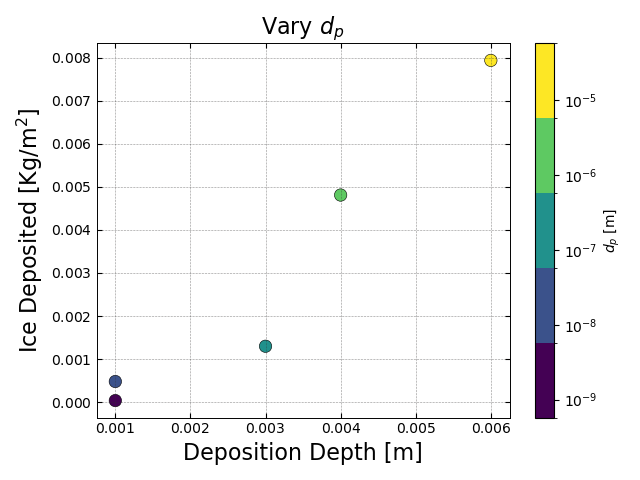}
 \caption{In this plot we show the total ice deposition (on the y-axis) and the final depth of the deposition front (on the x-axis) for a range of pore diameters, $d_p$ (shown in color with color bar on the right). The runs plotted are Runs 7-11 which parameters listed in Table \ref{tab:Runs}. We find that for pore diameters smaller than $d_p = 10^{-8}$ m, there is minimal ice deposition. 
 } \label{fig:vary_pore_radius_H2O_icy}
\end{figure}

We run a suite of simulations varying the pore diameter (See Figure \ref{fig:vary_pore_radius_H2O_icy} and Table \ref{tab:Runs}, Runs 7--11), tracking the depth of the start of the ice deposition front (the depth at which the volume ice fraction exceeds the initial ice fraction prior to the heating event) as well as the column density of deposited ices. This column density does not take into account mass which is lost to the environment, only the deposited ice which exceeds the initial ice fraction. We find that increasing the pore size increases both the depth of the deposition front as well as the quantity of ice deposited within the deposition front.

At small pore diameters, diffusion is less efficient and the deposition front is not able to penetrate deeply beneath the surface. Additionally, very little material is incorporated into the deposition front. As the pore diameter increases beyond $d_p = 10^{-8}$ m, ice deposition is more substantial as well as moves deeper beneath the surface.

\subsection{Sensitivity to Transient Heating Event Strength}

In this section, we investigate the effects of transient heating event strength on the volatile deposition front in an initially icy, porous body. The exact conditions of the runs can be found in Table \ref{tab:Runs} for Runs 2 (icy body), 4 ( $\Gamma = 10$), and 12-20. 
Here, Run 4 is similar to Run 2 but is an example of higher transient heating event strength, $\Gamma$. Runs 12-20 are similar to Run 2 but cover a range of $\Gamma$. We refer to a strong event as a high temperature in a protoplanetary disk. We do not have a detailed nebular shock model (e.g., \citealt{Ciesla_2002}); however, we approximate the effects of a shock by scaling the temperature and gas number density by a multiplier, $\Gamma$, and the pressure by the square of the multiplier, $\Gamma^2$. The $\Gamma$ required for deposition fronts to occur depend on the ambient environmental conditions, requiring the max temperature, pressure, and density to support gaseous volatiles. 

The fiducial transient heating event strength, $\Gamma = 5$, demonstrated in Figure \ref{fig:Icy_Body}, increases the nebular temperature and gas number density by 5 times and pressure by 25 times those of the ambient conditions. We simulate a range of values from $\Gamma = $ 2 to 10, with specific run conditions listed in Table \ref{tab:Runs}. 

\begin{figure}[!t]
 \centering
 \includegraphics[width=9cm]{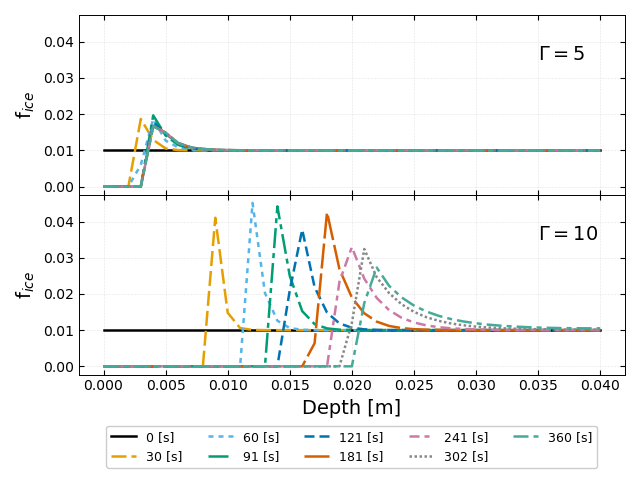}
 \caption{We plot the ice fraction as a function of depth and time throughout the simulation for Runs 2 and 4. Initially icy grains with different transient heating event strengths, $\Gamma = 5$ (Top) and $\Gamma = 10$ (Bottom). We find that increasing $\Gamma$ allows for heat, and subsequently the deposition front, to travel deeper beneath the surface of the body. For an initially icy body, the peak ice volume fraction is greater due to more native ice being incorporated in the deposition front.}
 \label{fig:Ice_Frac_Shock_Strength}
\end{figure}

Figure \ref{fig:Ice_Frac_Shock_Strength} shows the ice fraction for Runs 2 and 4. We find that weaker transient heating events lead to less deposited ice, along with the ice that is deposited being located closer to the surface. This is expected because a majority of the deposition front is associated with the native ices. As such, lower transient heating event strengths lead to less heating of the subsurface material, less sublimation, and subsequently less internal deposition. The opposite effect is seen for higher strength events. 

\begin{figure}[!t]
 \centering
 \includegraphics[width=9cm]{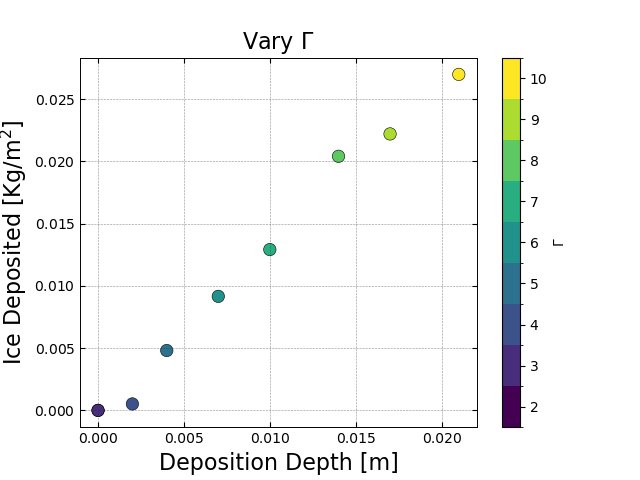}
 \caption{We plot the depth at which the deposition front begins versus the total ice deposited within the deposition front at the end of the simulation (t = 360 sec) for a range of transient heating event strengths, Runs 12-20 with parameters listed in Table \ref{tab:Runs}. 
 We denote the event strength as $\Gamma$, where $\Gamma = 5$ indicates a transient heating event with peak conditions 5 times that of the ambient environmental temperature and pressure. We hold all other variables constant while varying $\Gamma$, with the fiducial values for non-varying parameters given in Table \ref{tab:Fiducial}. Given our fiducial values, transient heating events with $\Gamma \le 3$ result in negligible deposition. The $\Gamma$ required for deposition fronts to occur depend on the ambient environmental conditions, requiring the max temperature, pressure, and density to support gaseous volatiles. Stronger transient heating events result in volatile deposition fronts to be deeper below the surface and have accumulated more material.
 }
 \label{fig:vary_shock_H2O_Icy}
\end{figure}

Figure \ref{fig:vary_shock_H2O_Icy} shows the total ice in the deposition front as a function of deposition depth for 8 simulations that differ in $\Gamma$. As more of the body is heated, more volatiles sublimate and travel deeper beneath the surface.

\subsection{Sensitivity to Transient Event Decay Timescale}
\label{SecDecay_Timescale}

\begin{figure}[!t]
 \centering
 \includegraphics[width=9cm]{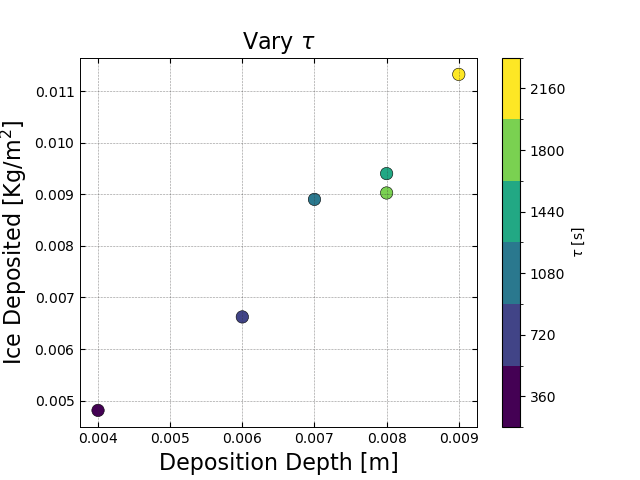}
 \caption{We plot the depth at which the deposition front begins versus the total ice deposited within the deposition front at the end of the simulation (t = 360 sec) for a range of decay timescales, Runs 21-26 with parameters listed in Table \ref{tab:Runs}. 
 The decay timescale, $\tau$, determines the decay rate as shown in Equation \ref{eq:DecayRate}. We hold all other variables constant while varying the transient heating event strength, with the fiducial values for non-varying parameters given in Table \ref{tab:Fiducial}. Longer duration heating events result in deeper volatile deposition fronts with more accumulated material.}
 \label{fig:vary_tau}
\end{figure}

In this section, we vary the decay timescale, $\tau$, for our simulations. We show the simulation parameters for these runs in Table \ref{tab:Runs}. We vary the decay timescale from 360 sec up to 2160 sec while keeping our simulation duration at 360 sec. As we increase the decay timescale, Figure \ref{fig:vary_tau} shows the deposition front moves deeper below the surface before becoming stationary due to the environmental temperatures being elevated for a longer duration. The amount of material within the deposition front also increases with longer heating events due to more locally sublimated volatiles being incorporated into the deposition front. 




\subsection{Other Volatiles (CO, CO\texorpdfstring{$_2$}{})} 
\label{Sec:OtherIces}

We have found that water-ice sublimates and subsequently deposits within a porous body's surface. In this section, we simulate transient heating events for two common carbon-bearing molecules, CO and CO$_2$, the building blocks of organic molecules \citep{Aikawa_1999}. Here, we show that these volatiles undergo deposition and sublimation phenomena similar to those of H$_2$O but at different orbital radii.




Using JWST/NIRSpec to observe Solar System bodies, CO and CO$_2$ ices have been detected on KBO/TNO surfaces at orbital radii of 30 to 50 AU \citep{DePra_2025, Markwardt_2025, Cryan_2025}. H$_2$O and CO$_2$ ices with little to no CO ice have been detected on KBO surfaces at 40 AU \citep{Wong_2025}, on Saturnian moons at orbital radii of 10 AU \citep{Belyakov_2025, Brown_2025}, and on Jupiter's moons and Trojans at orbital radii of 5 AU \citep{Wong_2024, Sharkey_2025}. Outside our Solar System, cold CO and CO$_2$ have been observed within icy layers at different heights of protoplanetary disk around low-mass pre-main sequence stars (e.g. \citealt{Powell_2022}, \citealt{Sturm_2023A}, \citealt{Sturm_2023B}). 

Since ice-phase CO and CO$_2$ can survive at a wide range of orbital radii, we simulate CO and CO$_2$ at orbital radii based on their freeze-out temperatures, assuming the disk midplane temperature calculated using Equation \ref{eq:AmbientTemperature}. Similar to interplanetary H$_2$O ice, interplanetary CO$_2$ becomes an icy solid below its triple point at 200~K, corresponding to an orbital radius within the midplane of 1 AU. We choose 5 AU, similar to our choice for H$_2$O. 
CO in mixtures with water-ice and/or N$_2$, chemically due to desorption, tends to be stable at a narrower temperature range than its triple point (60 K) between 10 and 40 K (e.g. \citealt{Palumbo_1997}, \citealt{Oberg_2005}, \citealt{Visser_2009}, \citealt{Qi_2024}), which for 20 K corresponds to an orbital radius within the midplane of 50 AU. We choose to model CO at 55 AU to guarantee that all CO is initially frozen and chemically stable. 

We expect volatile species to behave independently of one another because molecule-molecule interactions are rarer compared to molecular-wall interactions for the relatively low partial pressures and gas number densities expected within small pores (see Section \ref{subsubsec:Knudsen_Regime}). If a volatile has a sublimation temperature near in value to the peak temperature caused by a nebular shock, then a given sublimated volatile may eventually cool at different depths for each resulting deposition front. Multiple deposition fronts can exist within a porous body at once, with each deposition front comprising a single volatile. Within a given simulation we only track a single volatile, since multiple deposition fronts within a body act nearly independently of one another.

\begin{figure}[t]
 \centering
 \includegraphics[width=9cm]{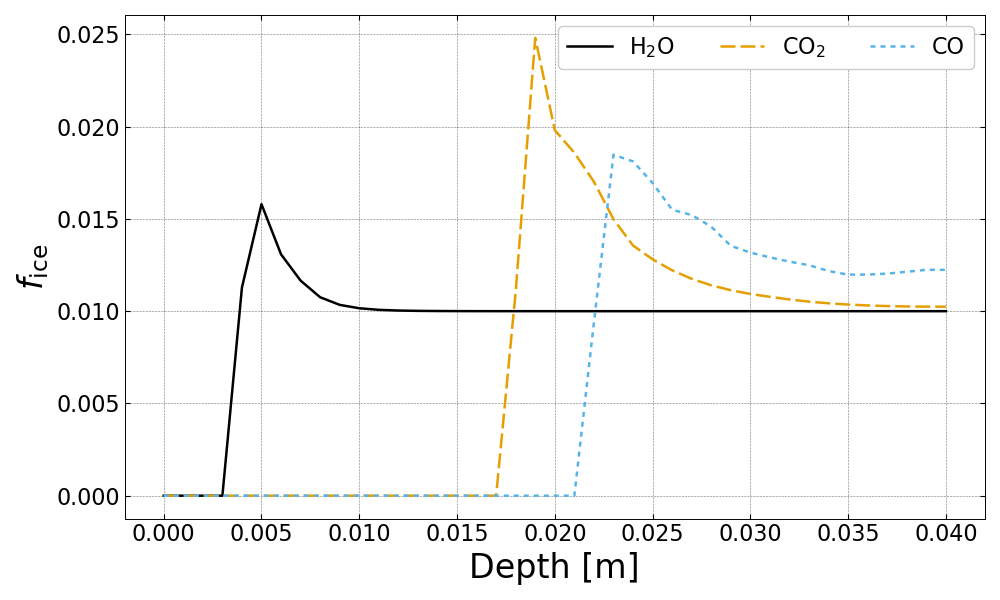}
 \caption{We plot the ice fraction as a function of depth at t = 360 sec for 3 different volatile ices: H$_2$O, CO$_2$, and CO (Runs 2, 5, and 6). The orbital radius and volatile gas fraction. Environmental conditions (a$_o$ and gas fraction) are not the same between volatiles shown, with the parameters for each simulation listed in Table \ref{tab:Runs}. Since simulating each volatile at different orbital radii results in similar ice fractions deposited, ice deposition fronts can occur throughout the protoplanetary disk, mainly varying in depth. 
 }
 \label{fig:vary_ice_type}
\end{figure}

The physical properties for each volatile are listed in Table \ref{tab:MolecularComponents}. For each volatile we simulate, sublimation occurs at different combinations of gas temperature and total pressure. 
Deposition fronts of H$_2$O, CO$_2$, and CO, shown in Figure \ref{fig:vary_ice_type}, all have similar profile shapes as a function of depth with a spike in ice fraction followed by rapid decay.

\subsection{Sensitivity to Porosity}

\begin{figure*}[!t]
 \centering
 \includegraphics[width=18cm]{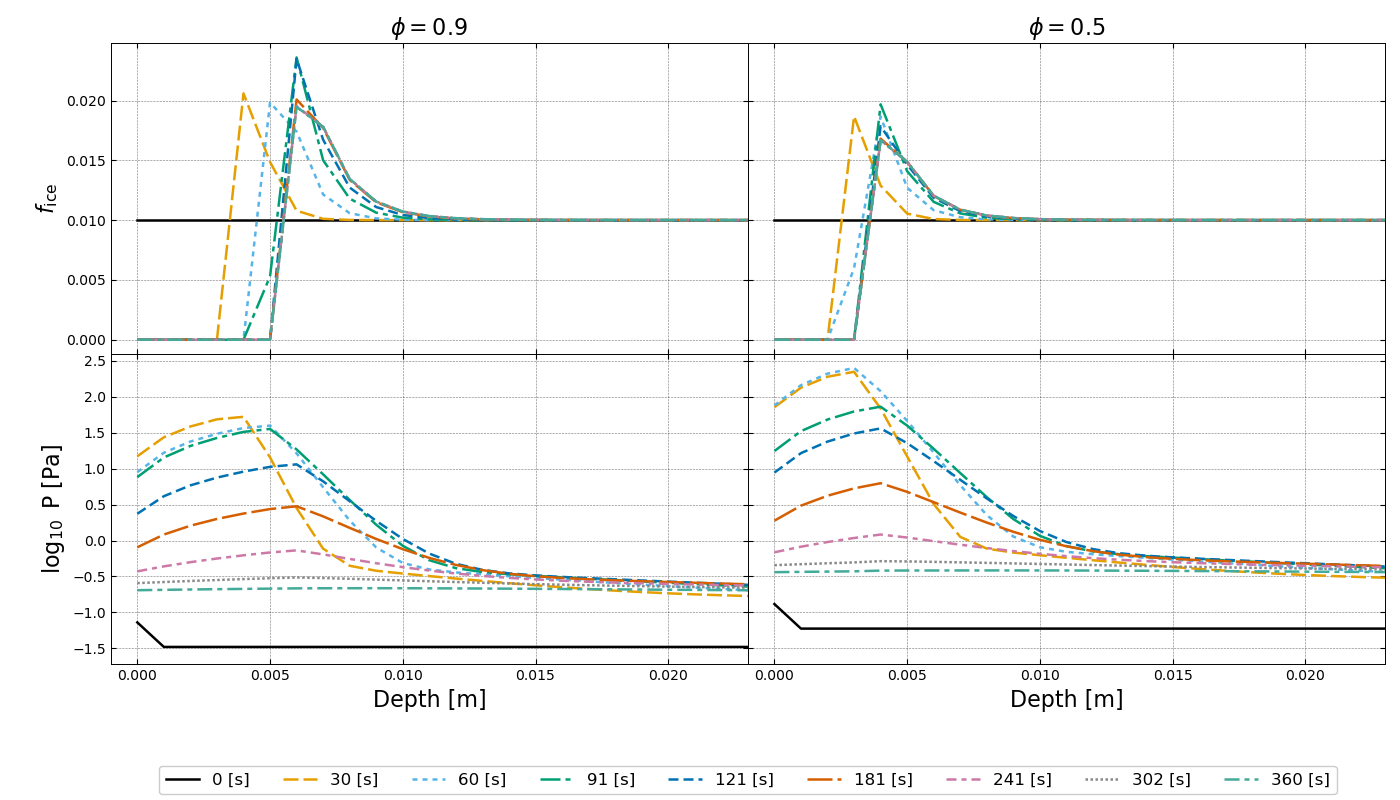}
 \caption{We plot the ice fraction and the internal pressure as a function of distance for Runs 27 and 35, an initially icy body with porosity of $\phi$ = 0.9 and $\phi$ = 0.5. This icy body is initially 1\% ice by volume. 
 \newline 
 Top: We find increasing the porosity results in a deposition front which penetrates deeper below the surface compared to a less porous body. With increased porosity, the deposition front occupies a greater fraction of solid material when compared to a less porous body. 
 We find increased porosity results in a more massive deposition front which penetrates deeper below the surface when compared to a less porous body in the same heating event. \newline
 Bottom: We find increased porosity results in lower peak pressures, with the pressure difference being nearly an order of magnitude between the extreme ends of the simulated range.}
 \label{fig:dual_porosity}
\end{figure*}

\begin{figure}
 \centering
 \includegraphics[width=9cm]{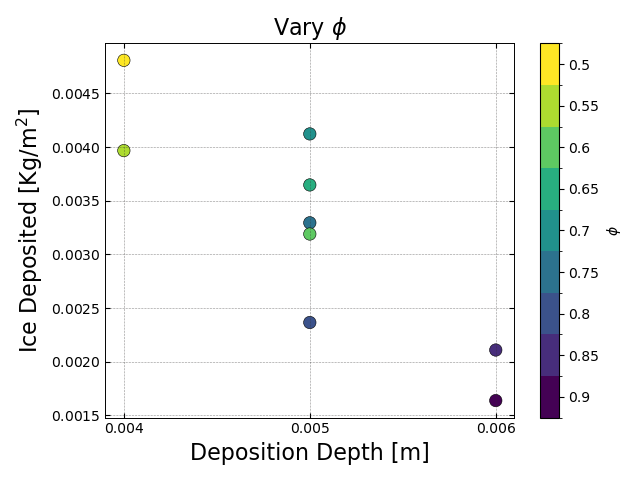}
 \caption{We plot the depth at which the deposition front begins vs the total ice deposited within the deposition front at the end of the simulation (t = 360 sec) for a range of porosities, Runs 27 - 35 in Table \ref{tab:Runs}. We find the depth of the deposition front is dependent on the porosity, however it is a weak dependency when compared to other tested parameters ($d_p$, $\Gamma$, $\tau$). }
 \label{fig:vary_porosity}
\end{figure}

We investigate the effects of varying the porosity, $\phi$, of the porous body from 0.9 to 0.1. The simulation conditions for these runs can be found in Table \ref{tab:Runs}, Runs 27 - 35. For porosity $\phi$ = 0.9 and $\phi$ = 0.5, we plot ice volume fraction $f_\text{ice}$ and pressure  in Figure \ref{fig:dual_porosity}. We find that as porosity increases, the deposition front penetrates deeper below the surface while simultaneously exhibiting lower peak pressures. With a porosity of $\phi = 0.5$, there is a small but noticeable decrease in the peak ice fraction as well as an order of magnitude increase in the peak pressure when compared to $\phi = 0.9$. 
As shown in Figure \ref{fig:vary_porosity}, the relatively lower peak ice fraction does not correlate with a lower total amount of ice deposition. More porous objects have less total solid material, resulting in higher ice fractions but less total ice deposited.

Changes in porosity have an analogous effect to changes in pore diameter on deposition front depth, peak ice fraction, and internal pressure. Increased porosity refers to an increase in the total void space within an object, while increased pore diameter refers to an increase in the void space contained within a single pore. This increase in void space, whether it be that contained within individual pores or the object as a whole, result in increased peak ice fractions and a decrease in the peak pressure.

\section{Discussion}
\subsection{
Longer Duration Heating Events}
\label{subsec:Long_Heating}



Chondrules form due to short-duration, hours-long nebular shocks set by the cooling time of the disk \citep{Desch_2002}, requiring cooling times of 100--1000 K/hr \citep{Yu_1995}. Our simulated disk undergoing a transient heating event with $\Gamma = 10$ and $\tau = 360$ sec would decay back to ambient conditions after 1 hour. This is an average cooling time of approximately 600 K/hr. Our simulation cools non-linearly. For the first 360 sec, our cooling rate is much more rapid with a variable slope on the order of 1000 K/hr. 


Short heating events with small decay timescales are also a proxy for larger porous bodies experiencing hours-long heating events. Due to gas drag, small particles move with the gas. However, objects exceeding the size-scale of meters are less well coupled to the gas. Any difference between gas and orbital velocity would allow a large object to pass through a shock. The duration of a transient heating event experienced by a large porous body could be shorter than that experienced by a mm-sized grain. While the decay rate in this simulation aligned with those of nebular shocks, the heating events associated with spiral density waves explored by \citet{Podolak_2011} could last for months, giving rise to long duration heating events for large decoupled bodies. Using Equation \ref{eq:skin_depth}, a month long heating event has a skin depth of $\sim 200$ m.  Large porous planetesimals would be capable of sequestering and enhancing the ice content below their surfaces during these long-term heating events.


\subsection{
Can Significant Amounts of Ice From Disk Volatiles be Deposited Within an Initially Dry Body?}
\label{sec:Discussion_Dry_Deposition}


In our initially dry body simulation of a transient heating event (Run 1, Figure \ref{fig:Dry_Body}), we found only small quantities of water vapor from the nebular gas was deposited as ice within the porous body. Increasing the initial ice content (See Run 2, Figure \ref{fig:Icy_Body}), we find the nebular volatile gas fraction does not strongly influence the ice deposition front because the volatiles are present primarily from local sublimation.
As the fraction of ice deposited is proportional to the volatile vapor pressure in the hot disk, a higher nebula gas density or a higher volatile gas fraction during the event could increase the amount of ice deposited, allowing for significant deposition fronts in dry bodies and external enrichment for deposition fronts in icy bodies.

Settling can increase the mass fraction of volatiles near the disk's midplane. 
We follow the work by \cite{Du_2014}, who model the protoplanetary disk's midplane warm water gas fraction between 1--1.2 AU using a gas fraction $X_{H_2O} \approx 10^{-4}$. We take the abundance of H$_2$O beyond the water ice line to be $X_{H_2O} \approx 10^{-4}$ \citep{Bosman_2018}. 
Note that we put a short decay timescale for the volatile vapor to cool in the simulation; however, it may take days or weeks for water vapor to reform into ice aggregates or condense onto rims of dust particles \citep{Sirono_2024}.  This deposition is primarily located on the surfaces of objects, indicating enrichment of environmental volatiles would be required for subsurface deposition fronts to be enhanced.

Being located in the disk midplane is insufficient to allow for significant subsurface deposition within the deposition front, thus we consider other methods of enhancing the volatile gas density. Spiral density waves can increase the local gas density by a factor of a few \citep{Dong_2015, Ziampras_2020, Cimerman_2021}. However, the peak ice fraction in Run 1 is so low that volatile gas densities would need to be multiple orders of magnitude higher than what we assumed in order to have a significant effect on the deposition quantity.
Another method to enhance ice deposition into initially dry bodies could be the collision of icy planetesimals \citep{Stewart_2025}. Collisions increase the density of water vapor into a jet (compared to the ambient level in the disk) by an order of magnitude.

In the nebular and bow shock models for chondrule formation, there is a variation in the gas velocity associated with the heating event. A recently formed chondrule can be destroyed by impacts from smaller particles with shorter stopping times \citep{Jacquet_2014}. A lower temperature heating event, if it involved acceleration of the gas, could also cause collisions between particles. Particles with different stopping times could experience varied collisions, and porous body surfaces could be processed due to collisions with smaller particles.
Collisions between small particles could be a source of nebular volatile vapor deposited inside a porous body if the porous body's interior acts as a cold sink. 

To summarize, even though spiral density waves and jets from icy planetesimal collisions can increase the local gas disk density, the ambient disk pressure and volatile fraction are so low that a 
short-duration transient heating event (of an hour or so) cannot significantly move ice from an icy grain reservoir into porous body centers during the event itself.

\subsection{Shielding and Transport}

We discuss the possibility of ice deposition fronts in shielding volatiles from future heating events and transporting volatiles throughout regions of a protoplanetary disk.

Ices that are deposited in the interior of a porous body could be shielded from subsequent short-term heating events. After an ice deposition front forms due to a heating event, a dry outer shell will form unless significant deposition occurs on the surface layer, as was discussed in Section \ref{sec:Discussion_Dry_Deposition}. We carried out simulations of sequential heating events, finding that a dry, porous outer shell heats quicker than an equivalent icy volume, allowing sequential heating events to continue moving ices deeper inside a porous body up until a deposition front is beyond the skin depth of a dry porous body for the encountered heating event. The icy body can then be transported to a different region of the disk or be accreted onto a planet or planetesimal, carrying with it ices sequestered beneath its surface, keeping volatile ices entrapped within less volatile ices \citep{Ligterink_2024, Williams_2025}, and sublimating volatiles as it is transported into the inner disk \citep{Krijt_2018, Schneider_2021}.


\subsection{CO Depletion and Enhancement}

We discuss the implications of ice deposition fronts in the context of CO enhancement and depletion within protoplanetary disks. 

Sequestering CO within H$_2$O and CO$_2$ ices is proposed within small porous bodies to retain CO in regions where sublimation occurs, depleting CO in the disk environment \citep{Williams_2025}. Host H$_2$O and CO$_2$ ices can entrap and enhance CO near the host ice's snow-lines \citep{Ligterink_2024, Williams_2025}. CO$_2$ ices may more efficiently entrap CO than H$_2$O ices can \citep{Simon_2019}. 

Host ices trapping CO could be extended to ice deposition fronts within larger porous bodies. Since CO$_2$ and CO ice deposition fronts occur deeper than an H$_2$O ice deposition front, if the native ice content is great enough, then the CO$_2$ deposition front could trap CO within closed pores. With multiple deposition fronts, CO could be sequestered by both H$_2$O and CO$_2$. CO would be unlikely to be entrapped within a newly forming deposition front due to CO sublimating at much lower temperatures than H$_2$O and CO$_2$. However, if an enclosed region forms, volatiles entrapped within this region would be shielded from heating events, allowing for entrapment and transport of materials throughout the disk in the interiors of planetesimals.

\subsection{Modeling Protoplanetary Disk Dust Grain Size Distributions}

Understanding how transient heating events alter the surface and subsurface morphology of porous pebbles and planetesimals may help to model the mass of porous dust grains distributed within protoplanetary disks.

\citet{Guerra_2024} modeled the Spectral Energy Distribution (SED) of HL Tau, a T Tauri star. The dust model for the SED took advantage of ALMA mm continuum images, which trace colder dust in the disk midplane and outer disk, to test the effects of porosity on dust mass estimates. They found porous grains result in a disk mass that is 3x greater than that found using compact grains.

\citet{Shi_2026} modeled the SED of MWC 480, a Herbig Ae/Be star. The dust emission model took advantage of ALMA mm continuum images and VLA cm continuum images, testing different grain size distributions and dust properties, including porosity. They found the compact grain model results in 3x to 6x the disk mass compared to that of the porous grain disk model. 

Pre-main sequence stars, according to models of ALMA and VLA images and SEDs, may contain meter-sized porous pebbles within 30--60 AU  
\citep{Zhang_2023, Shi_2026},  which may sequester ices during transient heating events (See Section \ref{sec:Results}).

Heating events which result in the loss of ices in small initially porous icy grains result in increased porosity for these grains, or increased surface porosity for larger bodies as a dry region is formed.
\citet{Gonzalez_2026} find porous grains may elevate dust-gas mass ratios. Additionally, simulations by \citet{Michoulier_2024} find that taking into account the change in the fragmentation line due to the CO snowline results in an increased dust-gas mass ratio when compared to not simulating a CO snowline.

\subsection{High Pore Pressures}

We ask whether the gas pressure during a transient heating event could cause fractures or brittle or ductile material failure within a porous body. 

Observations of dust jets from Comet 19P/Borrelly by the Deep Space 1 mission inferred subsurface temperature and pressure close to water's triple point \citep{Yelle_2004}.
The primary proposed mechanism allowing liquid water to form in small Solar System bodies is due to short-lived, hot radionuclides that were incorporated into 5–-500 km sized bodies shortly after the formation of the solar nebula \citep{Wallis_1980}. 

Capillary suction due to microporosity enhances the physical strength of porous, wet materials \citep{Miles_2012} so that solid crusts prevent volatiles from escape, forcing volatile buildup inside comets \citep{Sheldon_2005,Miles_2012}.
If the liquid phase becomes supersaturated with gas, then a heat pulse can trigger the sudden release of dissolved gases, leading to an outburst. 

In Run 3 (see Figure \ref{fig:Icy_Pore_Radius}), pore pressures approach $10^4$ Pa, about 10 times higher than the triple point of water, suggesting that liquid water can condense within the porous body given appropriate temperatures, pore size, and initial ice fraction.

Tensile strength within asteroids and planetesimals inferred from porous collections of soda-lime glass \citep{Sakurai_2025}, or following a formulation for tensile strength of aggregate and porous non-aggregate monomers \citep{Kimura_2020, Kimura_2025}, predict strengths on the order of $10^4 - 10^5$ Pa for boulders and icy aggregates. We find smaller pore diameters, stronger heating events and thus rapid heating, or increased ice reservoirs can lead to pressures approaching the strength of the body (See Figure \ref{fig:Icy_Pore_Radius}), potentially causing an outburst that ejects material. Such an outburst would be similar to a cometary outburst \citep{Bischoff_2018} but caused by transient heating of a porous planetesimal in a circumstellar disk, rather than due to solar illumination. We carried out additional simulations with an initial ice volume fraction of 5\%, finding increased initial ice content results in higher pressure, which could point to liquid water or fragmentation. However, this high pressure approaches regimes where Knudsen diffusion is no longer a good approximation and advective transport becomes relevant in the presence of liquids, both of which are neglected as simplifications in our simulations.

\subsection{Implications for Aqueous Alteration}

Solutes, such as inorganic salts, can facilitate melting volatile ices within Solar System bodies of all sizes. Secondary minerals, such as iron sulfide (FeS) assemblages identified in the returned Comet 81P/Wild 2 samples could have been formed by low-temperature aqueous alteration \citep{Berger_2011}, though well-preserved phyllosilicates are absent. Alternatively, aqueously altered material could have been incorporated into the comet body during its formation rather than due to cometary activity after formation \citep{Ogliore_2023}. Fibrous mackinawite (composed of FeS), discovered in samples of the asteroid Bennu returned by the OSIRIS-REx mission \citep{Benner_2026}, is likely formed via precipitation from solution at temperatures up to 70$^\circ$ C. 

Laboratory experiments find that hydrated silicates rapidly form from sub-micrometer amorphous silicates at room temperature in mildly alkaline solution, giving an alternative scenario for aqueous alteration of interplanetary dust particles \citep{Nakamura-Messenger_2011}.
While a number of laboratory experiments and mechanisms have been proposed for aqueous alteration in small, short-lived deposits of water in primordial materials, so far we lack definitive signatures that such processes took place. 

We find that transient heating events can approach conditions allowing for liquid water to be present within porous icy bodies. Due to simplifications made regarding heat and mass transport (See Sections \ref{subsubsec:Knudsen_Regime}, \ref{subsubsec:Advective_heat}), we cannot predict the presence of liquid water. This is a topic of future work, further discussed in Section \ref{sec:Future_Work}.

\section{Summary}

In this paper, we have explored how a porous body that has a size greater than a few cm and is embedded in a protoplanetary disk could be affected by a transient event that heats the ambient disk gas. We use 1-dimensional time dependent energy and mass transport equations as a function of depth within the body that take into account Knudsen Diffusion, sublimation and deposition of volatiles, and heating of an inert porous substrate. We assume a short transient heating event of 360 seconds and discussion of longer duration heating events can be found in Section \ref{subsec:Long_Heating}. In the Knudsen diffusion limit, where the gas density is low, gas diffusion through the porous medium can be separately simulated for each volatile. For a volatile such as water vapor, our simulations show an advancing front consisting of sublimation on one side and deposition of ice on the other side. For short, 6 minute heating events, we find that the phenomenon of subsurface ice deposition, previously seen in comet simulations (e.g., \citealt{Kossacki_2015,Spohn_2015,Zivithal_2025}), is also likely to occur on 10 cm and greater sized porous bodies during transient heating events in the protoplanetary disk. 
 
We find the depth and amount of ice in a deposition front are strongly dependent on pore size, initial ice quantity, peak temperature, and duration of the heating event. Due to the low pressure in a circumstellar disk, in a solid body that is initially dry, the amount of nebular material deposited as ice is negligible, as illustrated in Figure \ref{fig:Dry_Body}. However, if the body is initially icy, then locally sublimated gas can accumulate in the advancing deposition front. For example, the deposition front in the simulation of an initially icy body shown in Figure \ref{fig:Icy_Body} has an ice volume fraction that is a few times higher than the initial ice fraction. 
We find that with a smaller pore diameter, lower porosity, lower peak temperature, and shorter decay timescale, the peak ice volume fraction within the deposition front decreases. This decrease in peak ice fraction often corresponds to a decrease in the column density of the deposited material, however, the opposite is true when decreasing porosity. Due to lower porosity increasing the total amount of rocky substrate, the peak ice fraction is lower while the column density of deposited material is greater (See Figures \ref{fig:dual_porosity}, \ref{fig:vary_porosity}). In all other cases, changes in peak ice fraction and column density of deposited material are correlated.

We find the peak internal pressure due to sublimation depends strongly on pore diameter and porosity. Decreasing the pore diameter leads to a larger peak pressure, with internal pressures exceeding $10^3$ Pa and approaching  $10^4$ Pa at a pore diameter of $d_p = 10^{-8}$ m. We find that decreasing porosity leads to an increase in the peak internal pressure, where the peak internal pressure differs by nearly an order of magnitude when comparing porous bodies with a porosity of  $\phi = 0.9$ and $\phi = 0.5$.

If the pore size is small, then the maximum subsurface pressure reached near the ice deposition layer could be high enough to fracture the material or condense liquid water. In the context of comets, high pressures could account for outbursts. Here, we find that high pressure in pores could have occurred in porous planetesimals due to transient heating in a protoplanetary disk. The presence of liquid water has been proposed in comets (e.g., \citealt{Sheldon_2005,Miles_2012}). We suggest that pockets of liquid water caused by transient heating, even if short-lived, might allow formation of iron sulfides or salts. Consequently, transient heating events could cause some aqueous alteration. Localized aqueous alteration associated with transient heating could occur in a low pressure environment that is external to the interiors of larger planetesimals and where aqueous alteration is usually assumed to occur \citep{Wallis_1980}.  

Our simulations of carbon monoxide and carbon dioxide (See Section \ref{Sec:OtherIces}) suggest that volatile ice deposition layers caused by heating events could occur on porous planetesimals throughout a protoplanetary disk. However, transient heating events would not affect all volatiles in the same way. For example, a weak transient event could cause CO to sublimate and be deposited in a porous body in the outer solar system, but water and carbon dioxide would be sublimated and deposited in higher temperature events that could have occurred in the asteroid belt. Due to their different vapor pressure temperature dependence, carbon dioxide ice and water ice would be deposited at different depths in a single body. 

We found that an initially dry body did not acquire much ice during a single short heating event from nebular water vapor. However, the disk is likely to retain volatile vapor until either it escapes the system or is eventually incorporated as ice on the surface of grains or larger bodies. After a transient heating event, the surface of a larger body could accumulate a layer of ice on its surface from water vapor that remains in the disk \citep{Sirono_2024}. After a series of heating events, a porous body could have more than one ice layer for a given volatile.

\subsection{Future work}
\label{sec:Future_Work}

Future studies can improve the numerical method and carry out simulations of longer duration events. To improve the numerical method, the integration can be taken to higher order, and we could take into account molecular collisions and integrate multiple volatile species simultaneously. Were we to allow pore size variations during the simulation, we could simulate bodies with initially higher ice fractions. 

We used a 1D model, but sublimation fronts need not be planar. Neutron imaging of vacuum freeze drying has revealed channels and fingering in the distribution of ice in porous media \citep{VorhauerHuget_2020}. \citet{VorhauerHuget_2020} attributed this structure to variations in the pore size distribution. These simulations may place constraints on porosity and tortuosity distributions inside primordial bodies, their subsurface structure, and differing evolutionary scenarios.

Some of our simulations displayed high pore pressure, suggesting pressures high enough to condense liquid water. 
Numerous studies have focused on aqueous alteration inside bodies that are large enough to have central pressures above the triple point of water. 
The possibility of short-lived pockets of liquid water is intriguing, particularly if in the future we could identify an associated chemical signature, such as a type of salt or sulfide. The liquid water pockets might uniquely form in a time that is short compared to the thousands or millions of years for aqueous alteration processes inside large bodies (greater than a few km) that have central pressures above water's triple point pressure. 

\vskip 1 truein
{\bf Acknowledgments: }
This material is based upon work supported by NASA, USA grant
80NSSC21K0143.

---------------------------------------------------------
\bibliographystyle{elsarticle-harv}
\bibliography{IceballHeating}

\appendix

\section{Numerical Model}
\label{App:Num_model}

Our system consists of two coupled diffusion (or heat) equations, one for energy and evolving temperature, and one for gas transport and evolving gas density. 
The diffusion coefficients are not constant as they are dependent both on depth and time. In addition, there are source terms to take into account ice sublimation or deposition. 

We use a 1-dimensional model of a porous solid, with physical quantities depending on time, $t$, and depth below surface, $x$. The surface is set at $x=0$, with increasing values of $x$ indicating increasing depths below the surface.

In each simulation we track molecular hydrogen along with a single volatile species gas species and its associated ice. We use 1-dimensional arrays for temperature $T$, porosity $\phi$, the bulk number density of molecular hydrogen $n_{g,0}$ and the volatile $n_{g,i}$ ($i>0$), and the number density of the associated ice form $n_{s,i}$. These arrays give these physical quantities as a function of depth. 
The body is assumed to be primarily composed of forsterite, a porous rocky and non-volatile solid with a density and thermal conductivity listed in Table \ref{tab:MolecularComponents}.

We simplify the energy transport equation (Equation \ref{eq:final_heat}), only taking into account energy from adsorption of a single volatile gas species:
\begin{equation}
 (1-\phi)C_{\text{uni} }\frac{\partial T}{\partial t} = \frac{\partial}{\partial x}\left( k_\text{T,eff} \frac{\partial T}{\partial x}\right) - E_\text{sub}\dot{n}_{\text{sub},i}.
\label{eq:final_heat_ap}
\end{equation}
%
The universal specific heat $ C_{\text{uni}}$ is a fixed value and given in Equation \ref{eq:rhocp_uni}. The bulk thermal conductivity $k_{T,\text{eff}}$ is computed with Equation \ref{eqn:k_eff} and is function of depth and time. 

We calculate the transport of a gas species via Knudsen diffusion (repeating Equation \ref{eq:mass_transport}),
\begin{equation}
 \frac{\partial n_{g,i}}{\partial t} = \frac{\partial}{\partial x}\left(D_{\text{Kn},i}\frac{\partial n_{g,i}}{\partial x}\right) + \dot{n}_{\text{sub},i}.
\label{eq:mass_transport_ap}
\end{equation}
When calculating parameters that rely on the volume which is physically occupied (within the pores), we scale by the local porosity, e.g, partial pressure $P_i = (n_{g,i}/\phi)~k_BT$. The diffusion coefficient is calculated using Equation \ref{eq:Knudsen_Diffusion} and is sensitive to depth and time. The ice sublimation or deposition rate, giving $\dot n_{\text{sub},i}$, is described by Equation \ref{eqn:subrate} and is also a function of depth and time. 

Equations \ref{eq:final_heat_ap} \& \ref{eq:mass_transport_ap} both contain a first order time derivative and a second order spatial derivative,
so they are both similar to the generic heat equation. 

We use an equally spaced grid in depth $x$ with grid spacing $\Delta x$.
The second order spatial partial derivatives are computed with a central difference approximation with truncation error $(\Delta x)^2$
\begin{align}
 \frac{\partial^2 u}{\partial x^2}
 \approx \frac{u_{j+1} - 2 u_j + u_{j-1}}{(\Delta x)^2},
\end{align}
where subscripts refer to spatial grid positions. For a heat equation with diffusion coefficient $D$
\begin{align}
\frac{\partial u}{\partial t} = D 
\frac{\partial^2 u}{\partial x^2} .
\end{align}
a forward time central-space (FTCS) scheme for numerical integration is 
\begin{align}
 u^{n+1}_j = u^n_j +D \frac{\Delta t}{\Delta x^2}
 \left( u_{j+1}^n - 2 u_j^n + u_{j-1}^n\right)
\end{align}
where superscripts referring to the time-step. 
The dimensionless parameter 
\begin{align}
 \alpha = D \frac{\Delta t}{\Delta x^2}
\end{align}
serves as a Courant number, and the numerical scheme is stable, according to von Neumann stability analysis, for a time-step that is sufficient small that 
\begin{align}
 \alpha < \frac{1}{2}.
\end{align}

For both heat and gas transport equations, we use a modified version of the forward time central-space (FTCS) scheme for approximating the solution to the heat equation. 

\subsection{Energy Equation}

We approximate the heat transport equation (Equation \ref{eq:final_heat_ap}) with 
\begin{align}
 (1-\phi_j^n)C_{uni}\frac{T^{n+1}_j - T^n_j}{\Delta t} &= k_{T,\text{eff}} \frac{T_{j+1}^n - 2 T_j^n + T_{j-1}^n}{(\Delta x)^2} \nonumber \\
 & \ \ \ \qquad + E_{\text{sub},i}\dot{n}^n_{j,\text{sub},i}
 \end{align}
where subscripts refer to spatial grid positions and superscripts refer to the time-step. This gives 
\begin{align}
 T^{n+1}_j &= T_j^n+\frac{\Delta t}{(1-\phi_i^n)C_{uni}}\bigg{(}k_{T,\text{eff}}\frac{T_{j+1}^n - 2 T_j^n + T_{j-1}^n}{(\Delta x)^2} & \nonumber \\
 & \qquad \qquad \qquad \qquad \qquad +\ E_{\text{sub},i}\dot{n}^n_{j,\text{sub},i}
 \bigg{)}. \label{eqn:Tup}
\end{align}

The energy transport equation has a heat diffusivity 
\begin{align} 
D_\text{heat} = \frac{k_{T,\text{eff}}}{(1-\phi) C_\text{uni}}.
\label{eqn:Dheat}
\end{align} 
The effective Courant number for the numerical scheme of Equation \ref{eqn:Tup} 
\begin{align}
 \alpha_u \equiv \frac{\Delta t}{(\Delta x)^2}D_\text{heat} = 
 \frac{\Delta t}{(\Delta x)^2}\frac{k_{T,\text{eff}}}{(1-\phi)C_{uni}}, \label{eqn:alpha_u}
\end{align}
where the subscript `$u$' denotes that this number is based on the energy transport equation. 

\subsection{Mass Transport Equation}

In the gas transport equation (Equation \ref{eq:mass_transport_ap}), we note that the Knudsen diffusion coefficient is dependent on depth $x$ through the thermal velocity, $V_{\text{th},i}$. 
Using Equation \ref{eq:Knudsen_Diffusion} for the diffusion coefficient and Equation \ref{eq:V_th} for the thermal velocity, the gas transport equation becomes 
\begin{align}
 \frac{\partial n_{g,i}}{\partial t} = \frac{d_p}{3}\sqrt{\frac{8k_B}{\pi m_i}}\frac{\partial }{\partial x}\left(\sqrt{T}\frac{\partial n_{g,i}}{\partial x}\right) + \dot{n}_{\text{sub},i}, 
\end{align}
where $\dot{n}_{\text{sub},i}$ is the sublimation rate of gas species `$i$'. 
We calculate a dimensionless number serving the role of a Courant number for the gas transport equation (Equation \ref{eq:mass_transport_numerical} but only taking into account diffusion),
\begin{align}
 \alpha_m \equiv &\frac{d_p}{3}\sqrt{\frac{8k_BT}{\pi m_i}} \frac{\Delta t}{(\Delta x)^2}.
 \label{eqn:alpha_m}
\end{align}

We split the update for the gas number density into two steps. 
We first calculate the gas transport neglecting the adsorption rate term, $\dot{n}_{\text{sub},i}$. We then add on a term to take into account sublimation or deposition. 

Taking into account diffusion alone and neglecting the sublimation source term, the mass transport equation is approximated by the finite difference equation 
\begin{align}
\frac{n^{a}_{g,i}-n^n_{g,i}}{\Delta t} &= \frac{d_p}{3}\sqrt{\frac{8k_B}{\pi m_i}} \times 
\nonumber \\
& \Bigg{[} \frac{\left(\sqrt{T^n_{j+1} }- \sqrt{T^n_{j+1}}\right)(n_{j+1,g,i}^n - n_{j-1,g,i}^n)}{2(\Delta x)^2} +\nonumber \\
& \sqrt{T_j^n}\left(\frac{n^n_{j+1,g,i} + n^n_{j-1,g,i}-2n^n_{j,g,i}}{(\Delta x)^2}\right)\Bigg{]} .
\end{align}
The superscript $a$ denotes the fact that we are carrying out the first part of the update. 
This gives the number density 
\begin{align}
n^{a}_{j,g,i} = n^n_{j,g,i}& + \frac{d_p}{3}\sqrt{\frac{8k_B}{\pi m_i}} \frac{\Delta t}{(\Delta x)^2}\times \nonumber \\
&\Bigg{[} \frac{\left(\sqrt{T^n_{j+1}} - \sqrt{T^n_{j-1}}\right)(n_{j+1,g,i}^n - n_{j-1,g,i}^n)}{2(\Delta x)^2} +\nonumber \\
& \sqrt{T_j^n}\left(\frac{n^n_{j+1,g,i} + n^n_{j-1,g,i}-2n^n_{j,g,i}}{(\Delta x)^2}\right)\Bigg{]}. \label{eq:mass_transport_numerical}
\end{align}

With time-step set by the effective Courant number for the gas transport diffusion equation, a significant fraction of the available gas can be deposited as ice during a single time-step if the density of the volatile gas is low. Similarly when the amount of ice is low, a significant fraction of it can be sublimated in a time-step. To reduce sensitivity to the grid spacing and time-step we we analytically solve for the volatile gas number density using the adsorption rate (Equation \ref{eqn:subrate}). We update the array containing number density of the volatile gas in two steps, first using Equation \ref{eq:mass_transport_numerical}. Then afterwards we adjust the number density to take into account sublimation or deposition as described below.


We repeat Equation \ref{eqn:subrate} for the deposition or adsorption rate, assuming an equality rather than the approximate relation previously written,
\begin{align}
\dot n_{\text{dep},i} 
&= V_{\text{th},i} (n_{g,i}-n_{v,i})\frac{\alpha_{\text{ads},i}}{4d_p} = -\dot n_\text{sub,i} 
\label{eqn:subrate2}.
\end{align}
With 
\begin{align}
\frac{d n_{g,i}}{dt} &= - \dot n_{\text{dep},i} \\
&= - V_{\text{th},i} (n_{g,i}-n_{v,i})\frac{\alpha_{\text{ads},i}}{4d_p}
\end{align}
we can find a time dependent solution 
\begin{align}
n_{g,i}(t) = n_{v,i} + (n_{g,i}(0) - n_{v,i} ) \exp\left( -\frac{ V_{\text{th},i}(T) \alpha_{\text{ads},i}}{4d_p} t\right)
\end{align}
in the approximation that the thermal velocity and vapor pressure are fixed (ignoring their temperature variations during the time-step). 
After time $\Delta t$
\begin{align}
n_{g,i}^{n+1} &= n_{v,i}^n + (n_{g,i}^a - n_{v,i}^n) \exp \left( - \frac{V_{\text{th},i}(T^n) \alpha_{\text{ads},i} }{4 d_p}\Delta t \right) . 
\label{eq:ap_n_rho_final}
\end{align}
The expression gives the second part of our update for the gas number density array. As the expression only depends on one specific spatial grid point, we have neglected the spatial grid index. 
Here $n_{g,i}^a$ is the number density we calculated in the first half of the update taking into account diffusion and that was computed in Equation \ref{eq:mass_transport_numerical}.

\subsection{Time-step choice}

We adopt a single time-step $\Delta t$ for an entire simulation. We compute the effective Courant numbers $\alpha_m$, $\alpha_u$ from equations \ref{eqn:alpha_u} and \ref{eqn:alpha_m} using the maximum temperature in the nebula during the transient heating event. 
In practice we find that the Courant number for gas transport, $\alpha_m$, is always smaller than the Courant number for energy transport $\alpha_u$. 
This is because the heat diffusivity 
exceeds the Knudsen diffusion coefficient,
\begin{align} D_\text{heat} > D_{\text{Kn},i}.\end{align} 
We set $\alpha_m = 1/2$ and using Equation \ref{eqn:alpha_m}, we solve for $\Delta t$. We use a constant time-step marginally smaller than this value to ensure stability for the entire simulation.


\subsection{Pressure, Porosity, and Solids}
\label{App:P_Phi_Solids}

At each time-step, we calculate the partial pressure of each gas species within a pore, the vapor pressure for a volatile at the current temperature, the change in solid volatile material, and the porosity. In calculating the partial pressure and vapor pressure we use Equations \ref{eqn:Pi} and \ref{eqn:Pvk2},
\begin{align}
 P_{i}^n =& k_B T^n\frac{n_{g,i}^n}{\phi^n} \label{eq:ap_partialPressure} \\
 P_{v,i}^n =& e^{A_{v,i}-B_{v,i}/T^n} \label{eq:ap_vapor_pressure}.
\end{align}
This is computed on each spatial grid position independently, so we have neglected the $j$ spatial index. 
The number density associated with the vapor pressure of gas species $i$
\begin{align}
n_{v,i}^n = \frac{\phi^n P_{v,i}^n}{ k_B T^n}.
\end{align}

Using Equations \ref{eq:ap_n_rho_final} and \ref{eq:mass_transport_numerical}, 
the number density of gas species $i$ that was sublimated during the time-step is $n_{g,i}^{n+1} - n_{g,i}^a$. 
We update the bulk number density of the associated solid species 
\begin{equation}
 n_{s,i}^{n+1}=n_{s,i}^{n}
 - (n^{n+1}_{g,i} - n_{g,i}^a)
 . \label{eqn:nupdate}
\end{equation}


The fractional volume occupied by the solid of species $i$ is $n_{s,i}/V_{i}$, where $V_i$ is the volume of a single molecule of species $i$,
\begin{equation}
 V_{i} = \frac{m_i}{\rho_{s,i}}.
\end{equation}
After computing the solid number density 
with Equation \ref{eqn:nupdate}, 
we update the porosity using the volume percentage occupied by solids,
\begin{equation}
 \phi^{n+1}=1-\sum_i(V_i n_{s,i}^{n+1}).
\end{equation}


\subsection{Initial conditions \& Boundary Conditions}

All arrays when initialized are independent of depth. The initial pressure, temperature, and number density arrays are set to match the ambient nebular conditions. The porosity as well as ice and rock fraction are initialized uniformly, independent of depth, and are variables that can be changed at the start of the simulation. The initial solid number density array is determined by the initial porosity and solid species fraction.

The lower boundary condition we set as a Neumann boundary with derivative equal to 0. If index $J$ is the right-most index, the boundary condition for temperature is defined as 
\begin{equation}
 T_{J}^{n+1} = T_{J-1}^{n+1}.
\end{equation}
This lower boundary condition applies to all all updated arrays. The surface boundary condition is set up using a ghost boundary cell. The values in this ghost cell are set to the time dependent current nebular conditions. The surface boundary cell is then evaluated the same as any internal cell, using this ghost cell. 

Implementation of the boundary conditions is the final set of computations made in a single time-step in the simulation. While the surface boundary condition depends on values from the current timestep, the lower (Neumann) boundary condition requires the interior values to have already been computed for the next timestep.



\section{Numerical Checks}
\label{sec:ap_Numerical_Checks}

We limit the amount of gas sublimated or the amount of ice deposited to ensure the change in gaseous number density does not exceed the available volatile in the solid or let the pressure fall below the vapor pressure. The total number of molecules in a grid cell sublimated during a timestep is limited to be at most the number of molecules in ice present in that same grid cell. Hence, Equation \ref{eqn:nupdate} is modified to ensure that $n_{s,i}^{n+1} $ can never become negative.

\begin{figure}
 \centering
\includegraphics[width=8cm]{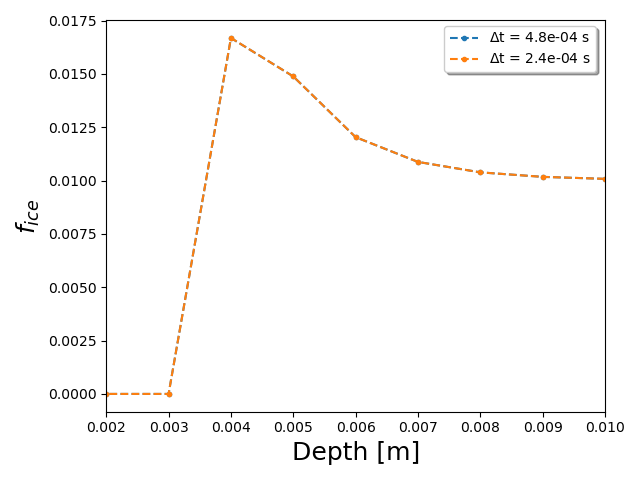}
 \caption{We plot the ice volume fraction as a function of depth at the end of the simulation for 2 different simulations. These simulations have the same parameters as Run 2 (with parameters listed in Table \ref{tab:Runs}) except for their time-step, $dt$, which is shown on the top right. We find that the estimated ice fractions are insensitive to time-step, with the deposition front being identical between both runs.}
 \label{fig:dt_variation}
\end{figure}

We have checked that our integrations are insensitive to simulation time-step duration, for time-steps shorter than that set from the effective Courant number $\alpha_m$.  This insensitivity is illustrated in Figure \ref{fig:dt_variation} where we show ice fractions as a function of depth for 2 different simulations that are the same except for their $\Delta t$ values.

\begin{figure}[!t]
    \centering
    \includegraphics[width=9cm]{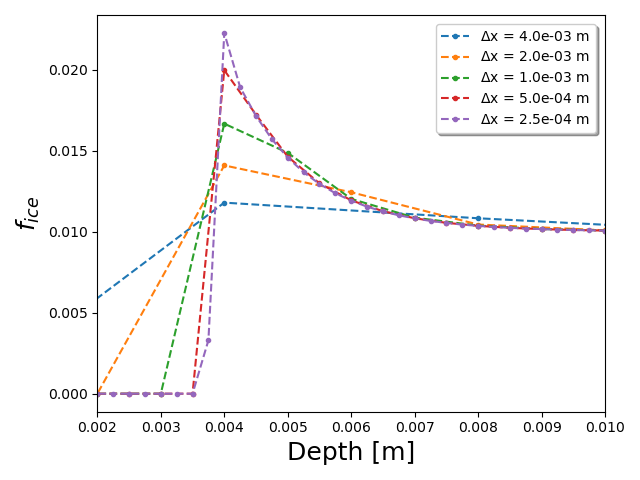}
    \caption{We plot the ice volume fraction as a function of depth and the end of the simulation ($t = 360$ s) for 5 different grid spacings. These simulations all have the same parameters as Run 2.}
    \label{fig:dx_variation}
\end{figure}

\begin{figure}[!t]
    \centering
    \includegraphics[width=9cm]{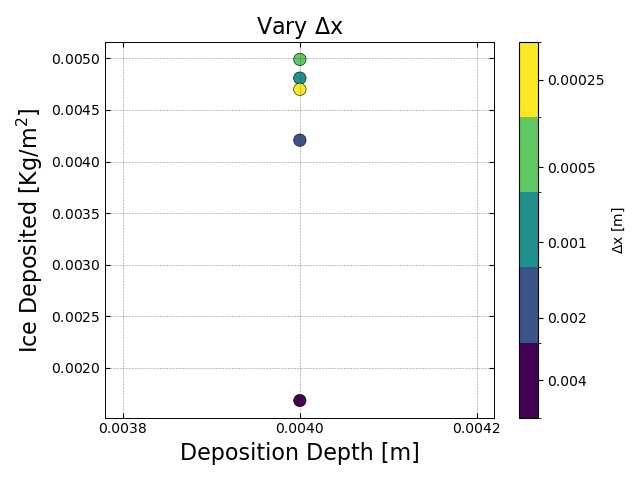}
    \caption{We plot the column density of deposited ices and the depth the deposition front exceeds the initial ice volume fraction for 5 different runs varying $\Delta x$. We find the column density converges toward a common value, approaching that convergence point with $\Delta x$ = 0.001 m. }
    \label{fig:dx_deposition}
\end{figure}

We additionally check that our integrations are insensitive to grid size. We plot in Figure \ref{fig:dx_variation} the ice volume fraction at $t = 360 s$ for 5 different grid spacings. By reducing the grid size, we find the deposition front becomes narrower and the ice volume fraction increases at the deposition front, which would be expected as volatiles primarily deposit at the earliest location possible, causing greater deposition in a smaller computational region when the grid size is reduced. In Figure \ref{fig:dx_deposition} we plot the column density of ice deposited and the depth at which the deposition front begins. We find the initial depth remains unchanged, also seen in Figure \ref{fig:dx_variation}, and the column density begins to converge with a grid spacing of $\Delta$ x = 0.001 m or smaller

\end{document}